\documentclass[aps,showpacs]{revtex4}

\def\ov#1{\overline{#1}}

\def\wt#1{\widetilde{#1}}
\def\vb#1{\mbox{\boldmath$#1$}}
\def\pd#1#2{\frac{\partial #1}{\partial #2}}

\def\wh#1{\widehat{#1}}
\def\bdot{\,\vb{\cdot}\,}
\def\btimes{\,\vb{\times}\,}

\def\bhat{\wh{{\sf b}}}
\def\cal#1{\mathcal{#1}}

\def\exd{{\sf d}}

\begin{document}

\title{Variational Principles for Reduced Plasma Physics}

\author{Alain J.~Brizard}
\address{Department of Physics, Saint Michael's College, Colchester, VT 05439, USA}

\begin{abstract}
Reduced equations that describe low-frequency plasma dynamics play an important role in our understanding of plasma behavior over long time scales. One of the oldest paradigms for reduced plasma dynamics involves the ponderomotive Hamiltonian formulation of the oscillation-center dynamics of charged particles (over slow space-time scales) in a weakly-nonuniform background plasma perturbed by an electromagnetic field with fast space-time scales. These reduced plasma equations are derived here by Lie-transform and variational methods for the case of a weakly-magnetized background plasma. In particular, both methods are used to derive explicit expressions for the ponderomotive polarization and magnetization, which appear in the oscillation-center Vlasov-Maxwell equations.
\end{abstract}


\maketitle

\section{\label{sec:intro}Introduction}

Our understanding of complex plasma phenomena can be significantly improved by the asymptotic elimination (or decoupling) of fast space-time scales from slow space-time scales of interest. Over the past 30 years, Lie-transform perturbation methods \cite{Kaufman_Lie,Cary_Lie,RGL_82} 
have played an important role in the process of dynamical reduction in plasma physics. Standard examples of reduced plasma dynamics include the elimination of fast space-time scales associated with the gyromotion of a charged particle in a strong magnetic field (i.e., the {\it guiding-center} reduction \cite{RGL_83,Kaufman_gc,CB_08} and the {\it gyrocenter} reduction \cite{KaufmanB_84,BH_07}) or the elimination of the fast eikonal space-time scales associated with charged-particle motion perturbed by a high-frequency, short-wavelength electromagnetic wave (the {\it oscillation-center} reduction \cite{JK_78,GKL,Cary_Kaufman,HW,Kaufman_Holm,SK,PLS,SKD,BMB,Fisch}).

Note that, while the process of dynamical reduction is implicitly associated with the elimination of fast space-time scales from a set of dynamical equations, the process may or may not explicitly reduce the number of degrees of freedom. When it does (e.g., guiding-center reduction), a fast orbital angle (used to describe particle motion) becomes asymptotically ignorable and its conjugate action becomes an adiabatic invariant. Whether or not the number of degrees of freedom is reduced, however, the numerical integration of these dynamically-reduced equations of motion over long space-time scales of interest, which generally allows for more realistic plasma geometries to be considered, presents an important practical application in plasma physics.

Variational formulations of reduced plasma dynamics exist for single-particle Hamiltonian dynamics as well as Vlasov-Maxwell (kinetic) and fluid self-consistent theories. First, the variational formulation for reduced single-particle dynamics relies on the dynamical reduction of the phase-space Lagrangian by Lie-transform methods \cite{RGL_82}. The reduced single-particle dynamics is expressed in terms of a reduced Hamiltonian and a reduced Poisson-bracket structure (derived from the symplectic part of the reduced phase-space Lagrangian). Conservation laws are associated with the invariance of the reduced Hamiltonian on fast degrees of freedom (e.g., ignorable or fast angles) and the corresponding invariants (e.g., exact or adiabatic actions) appear explicitly in the symplectic part of the reduced phase-space Lagrangian. Second, the variational formulation of the reduced Vlasov-Maxwell (kinetic) equations, on the other hand, is intimately connected to the variational formulation of reduced single-particle dynamics through a constrained variational principle in extended phase space \cite{Brizard_2000a}. Conservation laws for the reduced Vlasov-Maxwell equations are derived by applying the Noether method on the reduced Lagrangian density. Third, the variational formulation for reduced fluid dynamics relies on the existence of the reduced fluid Lagrangian density, from which exact conservation laws are derive once again by the Noether method
\cite{Brizard_NFLR}.

\subsection{Berkeley School of Plasma Physics}

In all of these developments, the Berkeley School of Plasma Physics, led by Allan N.~Kaufman in collaboration with his graduate students and postdocs, has played a fundamental role. Allan's pioneering work in plasma theory over 50 years (with over 100 papers), which is reviewed in his memoirs (to appear in the KaufmanFest Proceedings), can be divided into three separate periods. During the first period (from 1957 to 1969), Allan's work focussed on plasma dynamics in a magnetic field (in collaborations with S.~Chandrasekhar and K.~M.~Watson) and the statistical physics of an imperfect gas. During the second period (from 1970 to 1987), Allan's work focussed on nonlinear plasma waves, the Hamiltonian formulation of plasma physics (which included the Lie-transform perturbation analysis of oscillation-center and guiding-center dynamics), the construction of dissipative Poisson brackets (by combining an entropy-conserving Poisson bracket and an energy-conserving dissipative bracket), wave \& particle chaos, and wave kinetic theory. During the third period (from 1987 to the present), Allan's work focussed on linear wave conversion in plasmas and fluids, as well as the non-eikonal formulation of wave-action conservation laws in plasma physics. 

The main characteristics of the Berkeley School of plasma physics, which were developed mainly during the second and third periods of Allan's work, involve the development and applications of Hamiltonian and Lagrangian methods in plasma physics. In fact, with a few exceptions (e.g., R.~L.~Dewar 
\cite{Dewar}, P.~J.~Morrison \cite{PJM}, and J.~A.~Krommes \cite{JAK}), the use of these methods in plasma physics is intimately associated with Allan's name. It is worth mentioning that the Berkeley School has also greatly benefited from the influx of many postdocs (from Princeton's PPPL and elsewhere), visitors, and collaborators (see Allan's memoirs for details).

The development and applications of Hamiltonian and Lagrangian methods in linear wave conversion (during the third period) are reviewed by Tracy and Brizard in a separate paper in the KaufmanFest Proceedings. Here, I present the derivation of the oscillation-center dynamics of charged particles (developed during the second period) in a weakly-magnetized background plasma perturbed by high-frequency electromagnetic field fluctuations. The emphasis will be placed on applications of Lie-transform perturbation methods (with John Cary \cite{Cary_Lie} and Robert Littlejohn \cite{RGL_82}, two of Allan's graduate students, acting as major participants in their developments) and applications of variational (action) principles to the development of the ponderomotive Hamiltonian oscillation-center theory (with Allan's students John Cary \cite{Cary_Kaufman} and Bruce Boghosian \cite{BMB} as well as a series of Allan's postdocs including Shayne Johnston \cite{JK_78}, Celso Grebogi \cite{GKL}, and Philippe Similon \cite{SK,PLS,SKD} 
acting as major participants in their developments). The purpose of the paper is not to present a complete survey of the historical development of oscillation-center Hamiltonian dynamics in plasma physics (which is briefly discussed in Allan's memoirs). Instead I wish to present a new (and hopefully interesting) application of Lie-transform perturbation and variational methods in plasma physics.

In the present paper, the ponderomotive (eikonal-averaged) polarization and magnetization effects due to a high-frequency electromagnetic wave propagating in a weakly-magnetized plasma are investigated by two complementary methods. The ponderomotive polarization and magnetization are derived either from an oscillation-center variational principle involving the weak background electromagnetic fields $({\bf E}_{0},{\bf B}_{0})$ as variational fields, or from the (Lie-transform) push-forward relation between the particle fluid moments and the oscillation-center fluid moments. An important consequence of the variational formulation is that an exact energy-momentum conservation law can be derived by the Noether method and is expressed in terms of linear and ponderomotive polarization and magnetization. We note that while some of the fundamental issues raised by this work are resolved within a covariant relativistic treatment, we present a non-relativistic treatment here (except when discussing the moving-magnetic-dipole controbution to polarization in Sec.~\ref{subsec:pf_pond}) and postpone the relativistic treatment for future work.

\subsection{Organization}

The remainder of the paper is organized as follows. In Sec.~\ref{sec:exact}, the variational formulation of the exact Vlasov-Maxwell equations is reviewed \cite{Brizard_2000a}. The formulation is given in terms of extended phase-space coordinates, in which the energy-time canonical-coordinate pair is added to the six-dimensional regular phase-space coordinates. From the Vlasov-Maxwell action functional, we also derive conservation laws by using the Noether method. In Sec.~\ref{sec:reduced}, the variational formulation of a generic set of reduced Vlasov-Maxwell equations is presented
\cite{Brizard_08}. Here, the terms ``exact'' and ``reduced'' are used to emphasize the fact that, while the exact Vlasov-Maxwell equations exhibit the full range of space-time scales associated with exact particle Hamiltonian dynamics, the reduced Vlasov-Maxwell equations exhibit only the long space-time scales associated with reduced Hamiltonian dynamics. The explicit derivations of general expressions for the polarization and magnetization by push-forward (Lie-transform) method and by variational method are presented. The energy-momentum conservation laws for the self-consistent reduced Vlasov-Maxwell equations are also derived by the Noether method. In Sec.~\ref{sec:Lie}, the Lie-transform perturbation analysis of oscillation-center Hamiltonian dynamics is presented in terms of non-canonical coordinates, so that both the Poisson-bracket (symplectic) structure and the Hamiltonian are perturbed by the electromagnetic wave fields. Next, the linear and nonlinear polarization and magnetization are derived by variational and push-forward methods. Lastly, a summary of this work is presented in Sec.~\ref{sec:sum}. 

\section{\label{sec:exact}Exact Vlasov-Maxwell Equations}

\subsection{Extended Hamiltonian Dynamics}

The most general setting for Hamiltonian perturbation theory \cite{Brizard_08} is the eight-dimensional extended phase space with coordinates $z^{a} = 
(t, {\bf x}; w, {\bf p})$, where $w$ and ${\bf p}$ denote the kinetic energy-momentum coordinates and $({\bf x}, t)$ denote the space-time location of a charged particle (of mass $m$ and charge $e$). Here, the Hamiltonian description of charged-particle dynamics in an electromagnetic field ${\sf F}_{\mu\nu} \equiv \partial_{\mu}A_{\nu} - \partial_{\nu}A_{\mu}$, represented by the four-potential $A_{\mu} \equiv (-\,\Phi,
{\bf A})$ [using the space-time metric $g^{\mu\nu} = {\rm diag}(-1, 1, 1, 1)$], is expressed in extended phase space in terms of (i) the extended Hamiltonian
\begin{equation}
\cal{H}({\sf z}) \;=\; \frac{1}{2m}\,|{\bf p}|^{2} \;-\; w \;\equiv\; 0,
\label{eq:extHam_def}
\end{equation}
and (ii) the extended phase-space Lagrangian (summation over repeated indices is implied)
\begin{equation}
\Gamma \;=\; \left( {\bf p} \;+\; \frac{e}{c}\,{\bf A}\right)\bdot \exd{\bf x} \;-\; \left( w \;+\; e\,\Phi\right)\;\exd t \;=\; 
\left( p_{\mu} \;+\; \frac{e}{c}\,A_{\mu}\right)\;\exd x^{\mu} \;\equiv\; \Gamma_{a}\;\exd z^{a},
\label{eq:extPSL_def}
\end{equation}
where $x^{\mu} \equiv (ct, {\bf x})$ and $p_{\mu} \equiv (-\,w/c, {\bf p})$. The constraint ${\cal H} = 0$ in Eq.~(\ref{eq:extHam_def}) implies that the physical motion takes place on the surface $w = |{\bf p}|^{2}/2m$.

The extended equations of motion are obtained from the variational principle
\begin{equation}
0 \;=\; \delta\;\left.\left.\int \right( \Gamma \;-\; {\cal H}\;d\tau \right),
\label{eq:ext_EL_VP}
\end{equation}
where $\tau$ is the Hamiltonian orbit parameter in extended phase space. The variational principle (\ref{eq:ext_EL_VP}) yields the Euler-Lagrange equations
\begin{equation}
\omega_{ab}\;\frac{dz^{b}}{d\tau} \;=\; \pd{{\cal H}}{z^{a}},
\label{eq:ext_EL}
\end{equation}
where the Lagrange matrix $\vb{\omega}$ (with components $\omega_{ab} \equiv \partial_{a}\Gamma_{b} - \partial_{b}\Gamma_{a}$) is associated with the differential two-form 
\begin{equation} 
\omega \;=\; \exd\Gamma  = \exd p_{\mu}\;\wedge\;\exd x^{\mu} \;+\; \frac{e}{2c}\,{\sf F}_{\mu\nu}\;\exd x^{\mu}\;\wedge\;\exd x^{\nu}
\;\equiv\; \frac{1}{2}\;\omega_{ab}\;\exd z^{a}\;\wedge\;\exd z^{b}.
\label{eq:omega_def}
\end{equation}

The extended Poisson bracket $\{\;,\;\}$ is obtained from the extended phase-space Lagrangian (\ref{eq:extPSL_def}) by inverting the Lagrange matrix 
(with components $\omega_{ab}$) to obtain the Poisson matrix ${\bf J} \equiv \vb{\omega}^{-1}$. From the Poisson-matrix components $J^{ab} \equiv \{ z^{a},\; z^{b}\}$, we obtain the extended Poisson bracket 
\begin{equation}
\{ F,\; G \} \;=\; \left( \pd{F}{x^{\mu}}\;\pd{G}{p_{\mu}} \;-\; \pd{F}{p_{\mu}}\;\pd{G}{x^{\mu}} \right) \;+\;
\frac{e}{c}\;{\sf F}_{\mu\nu}\;\pd{F}{p_{\mu}}\,\pd{G}{p_{\nu}} \;\equiv\; \pd{F}{z^{a}}\;J^{ab}({\sf z})\;\pd{G}{z^{b}},
\label{eq:extPB_def}
\end{equation}
defined in terms of two arbitrary functions $F$ and $G$. Hamilton's equations in extended phase space are expressed as
\begin{equation}
\frac{dz^{a}}{d\tau} \;=\; \left\{ z^{a},\; \cal{H}\right\} \;=\; J^{ab}({\sf z})\;\pd{\cal{H}({\sf z})}{z^{b}},
\label{eq:extHameq_def}
\end{equation}
which includes 
\begin{equation}
\frac{dx^{\mu}}{dt} \;=\; \pd{{\cal H}}{p_{\mu}} \;\equiv\; v^{\mu} \;=\; \left( c,\; {\bf v} = \frac{{\bf p}}{m} \right),
\label{eq:dxmu_dt}
\end{equation} 
and $dp_{\mu}/dt = (e/c)\;{\sf F}_{\mu\nu}\,v^{\nu}$, when $dt/d\tau = 1$ is substituted.

Next, the extended Vlasov equation, which describes the time evolution of the particle distribution on extended phase space, is expressed in terms of the extended Hamiltonian (\ref{eq:extHam_def}) and extended Poisson bracket 
(\ref{eq:extPB_def}) as
\begin{equation}
0 \;=\; \frac{d\cal{F}}{d\tau} \;=\; \frac{dz^{a}}{d\tau}\;\pd{\cal{F}({\sf z})}{z^{a}} \;\equiv\;  \{ \cal{F},\; \cal{H}\},
\label{eq:extVlasoveq_def}
\end{equation}
where, in order to satisfy the physical constraint (\ref{eq:extHam_def}), the extended Vlasov distribution is defined as
\begin{equation}
\cal{F}({\sf z}) \;\equiv\; c\,\delta(w - |{\bf p}|^{2}/2m)\; f({\bf x}, {\bf p}, t),
\label{eq:F_extended} 
\end{equation}
and $f({\bf x}, {\bf p}, t)$ denotes the time-dependent Vlasov distribution on regular phase space. By integrating the extended Vlasov equation 
(\ref{eq:extVlasoveq_def}) over the energy coordinate $w$ (and using $d\tau = dt$), we obtain the regular Vlasov equation
\begin{equation} 
0 \;=\; \frac{df}{dt} \;\equiv\; \pd{f}{t} \;+\; \frac{d{\bf x}}{dt}\bdot\pd{f}{{\bf x}} \;+\; \frac{d{\bf p}}{dt}\bdot\pd{f}{{\bf p}}.
\label{eq:Vlasov_def}
\end{equation}
Lastly, the extended Vlasov equation (\ref{eq:extVlasoveq_def}) is coupled to Maxwell's equations
\begin{eqnarray}
\nabla\bdot{\bf E} & = & 4\pi\,\rho, \label{eq:E_Poisson} \\
\nabla\btimes{\bf B} - \frac{1}{c}\pd{{\bf E}}{t} & = & \frac{4\pi}{c}\,{\bf J}, \label{eq:B_Ampere}
\end{eqnarray}
where ${\bf E} \equiv -\,\nabla\Phi - c^{-1}\partial{\bf A}/\partial t$ and ${\bf B} \equiv \nabla\btimes{\bf A}$ satisfy the constraints $\nabla\bdot
{\bf B} = 0$ and $\nabla\btimes{\bf E} + c^{-1}\,\partial_{t}{\bf B} = 0$, and the charge-current densities
\begin{equation} 
\left( \begin{array}{c}
\rho \\
{\bf J}
\end{array} \right) \;=\; \sum\;e\;\int d^{4}p\;\cal{F}\;
\left( \begin{array}{c}
1 \\
{\bf v}
\end{array} \right) \;=\; \sum\;e\;\int d^{3}p\;f\;\left( \begin{array}{c}
1 \\
{\bf v}
\end{array} \right)
\label{eq:rhoJ_def}
\end{equation}
are defined in terms of the extended Vlasov distribution (\ref{eq:F_extended}), with $d^{4}p = c^{-1}dw\,d^{3}p$.

\subsection{\label{subsec:VP_exact}Vlasov-Maxwell variational principle}

The Vlasov-Maxwell equations (\ref{eq:extVlasoveq_def}) and (\ref{eq:E_Poisson})-(\ref{eq:B_Ampere}) can be derived from the Vlasov-Maxwell action functional \cite{Brizard_2000a}
\begin{equation}
{\cal A}[{\cal F},A_{\mu}] \;=\; -\;\sum\;\int d^{8}z\,{\cal F}({\sf z}) \;{\cal H}({\sf z}) \;+\; \int \frac{d^{4}x}{16\,\pi} \;{\sf F}:{\sf F}
\;\equiv\; \int d^{4}x\;{\cal L}(x).
\label{eq:Action_exact}
\end{equation}
The variation of the Lagrangian density
\begin{equation}
{\cal L}(x) \;\equiv\; \frac{1}{16\pi}\;{\sf F}(x):{\sf F}(x) \;-\; \sum\;\int d^{4}p\;{\cal F}(x,p)\; {\cal H}(p)
\label{eq:Lag_exact}
\end{equation}
yields (after rearranging terms)
\begin{equation}
\delta{\cal L} \;\equiv\; \delta A_{\mu} \left(\; \frac{1}{4\pi}\,\pd{{\sf F}^{\nu\mu}}{x^{\nu}} \;+\; \sum\;\frac{e}{c} \int  d^{4}p\; 
{\cal F} \;\pd{{\cal H}}{p_{\mu}} \;\right) \;-\; \sum\;\int  d^{4}p\;\; {\cal S}\, \{ {\cal F},\;{\cal H}\} \;+\; \partial_{\nu}\;\Lambda^{\nu}.
\label{eq:deltaL_exact}
\end{equation}
To obtain this expression, we used the constrained (Eulerian) variation for ${\cal F}$, defined as
\begin{equation}
\delta{\cal F} \;\equiv\; \Delta\cal{F} \;-\; \delta z^{a}\;\partial_{a}\cal{F} \;=\; -\; \delta z^{a}\;
\partial_{a}\cal{F},
\label{eq:deltaF_def}
\end{equation}
where the Lagrangian variation $\Delta\cal{F} \equiv 0$ (by definition since the Vlasov distribution is constant along a {\it Lagrangian} orbit in phase space). The virtual displacement in extended phase-space is 
\begin{equation}
\delta z^{a} \;\equiv\; -\; \left( \{ \cal{S},\; z^{a} \} \;+\; \frac{e}{c}\,\delta A_{\mu}\;\left\{ x^{\mu}, z^{a}\right\} \right), 
\label{eq:deltaz_def}
\end{equation}
where $\cal{S}$ is the canonical generating scalar field for this phase-space displacement. Hence, by substituting the displacement 
(\ref{eq:deltaz_def}) into Eq.~(\ref{eq:deltaF_def}), we obtain the Eulerian variation 
\begin{equation}
\delta{\cal F} \;\equiv\; \{ \cal{S},\; \cal{F} \} \;+\; \frac{e}{c}\,\delta A_{\mu}\;\pd{{\cal F}}{p_{\mu}}. 
\label{eq:deltaF_exact}
\end{equation}
Lastly, the last term in Eq.~(\ref{eq:deltaL_exact}) involves the Noether four-vector
\begin{equation} 
\Lambda^{\nu} \;\equiv\;  \frac{1}{4\pi}\;\delta A_{\mu}{\sf F}^{\mu\nu} \;+\; \sum\;\int d^{4}p\; {\cal S} \left( {\cal F}\;
\pd{{\cal H}}{p_{\nu}} \right),
\label{eq:Noether_exact}
\end{equation}
which does not contribute to the variational principle
\begin{equation}
\int\; \delta{\cal L}\; d^{4}x \;=\; 0
\label{eq:VP_exact}
\end{equation}
because this term appears as a space-time divergence in Eq.~(\ref{eq:deltaL_exact}). 

\subsubsection{Vlasov-Maxwell Equations.}

We obtain the extended Vlasov equation (\ref{eq:extVlasoveq_def}) by simply requiring stationarity in Eq.~(\ref{eq:VP_exact}) with respect to ${\cal S}$. Stationarity with respect to $\delta A_{\nu}$, on the other hand, yields the Maxwell equations
\begin{equation} 
\frac{1}{4\pi}\;\partial_{\nu}{\sf F}^{\nu\mu} \;=\; -\,\sum\;\frac{e}{c} \int d^{4}p\;
\pd{{\cal H}}{p_{\mu}}\,{\cal F} \;=\; -\,\sum\;\frac{e}{c} \int d^{3}p\;\;v^{\mu}\, f.
\label{eq:MaxVP_exact}
\end{equation}
Decomposition in terms of components of ${\sf F}$ yields Eqs.~(\ref{eq:E_Poisson})-(\ref{eq:B_Ampere}). Note that the electromagnetic field tensor also satisfies the Maxwell {\it constraint} equations $\partial_{\sigma}{\sf F}_{\mu\nu} + \partial_{\mu}{\sf F}_{\nu\sigma} + \partial_{\nu}
{\sf F}_{\sigma\mu} = 0$. 

\subsubsection{Energy-Momentum Conservation Law.}

An important advantage of a variational formulation is that one can use Noether's Theorem \cite{Brizard_05} to derive explicit conservation laws. The conservation laws for the Vlasov-Maxwell equations (\ref{eq:extVlasoveq_def}) and (\ref{eq:MaxVP_exact}) are derived from the Noether equation 
$\delta{\cal L} = \partial_{\nu}\Lambda^{\nu}$. By substituting explicit expressions for the variations $({\cal S}, \delta A_{\mu}, \delta{\cal L})$, we can obtain the conservation laws of energy-momentum, angular momentum, and wave action. The latter conservation law plays a crucial role in the analysis of linear wave conversion in plasmas (see paper by Tracy and Brizard in the KaufmanFest Proceedings). 

We now present the derivation of the energy-momentum conservation law associated with invariance of the Lagrangian density (\ref{eq:Lag_exact}) with respect to space-time translation $x^{\mu} \rightarrow x^{\mu} + \delta x^{\mu}$. Under this translation, the variations $({\cal S}, \delta A_{\mu}, \delta{\cal L})$ are 
\begin{equation}
\left. \begin{array}{rcl}
{\cal S} & \equiv & \left( p_{\mu} + eA_{\mu}/c\right)\,\delta x^{\mu}  \\
\delta A_{\mu} & \equiv & {\sf F}_{\mu\nu}\;\delta x^{\nu} - \partial_{\mu}\left( A_{\nu}\;\delta x^{\nu}\right) \\
\delta{\cal L} & \equiv & -\;\partial_{\mu}( \delta x^{\mu}\;{\cal L})
\end{array} \right\},
\label{eq:variations_exact}
\end{equation}
where ${\cal S}$ generates the virtual displacement $\delta x^{\mu} = \{ x^{\mu},\; {\cal S} \}$ as required. After substituting the variations (\ref{eq:variations_exact}) in the Noether equation (\ref{eq:Noether_exact}), and using the Maxwell equations (\ref{eq:MaxVP_exact}), we obtain the energy-momentum conservation law $\partial_{\mu}\,{\sf T}^{\mu\nu} \equiv 0$, where the stress-energy tensor 
\begin{equation}
{\sf T}^{\mu\nu} \;\equiv\; \frac{1}{4\pi} \left[\; \frac{1}{4}\,g^{\mu\nu}\;({\sf F}:{\sf F}) \;-\; ({\sf F}\cdot
{\sf F})^{\mu\nu} \;\right] \;+\; \sum\;\int d^{4}p\; \left( \pd{{\cal H}}{p_{\mu}}\,p^{\nu}\right)\;{\cal F}
\label{eq:EMCon_exact}
\end{equation}
is the sum of the field contribution (involving the tensor ${\sf F}$) and the particle Vlasov contribution (involving the Vlasov distribution ${\cal F}$).

\section{\label{sec:reduced}Reduced Plasma Dynamics}

The processes of dynamical reduction in single-particle plasma dynamics and plasma kinetic theory are associated with the extended near-identity phase-space transformation $\cal{T}_{\epsilon}: {\sf z} \rightarrow \wh{{\sf z}} = \cal{T}_{\epsilon}{\sf z}$, and its inverse $\cal{T}_{\epsilon}^{-1}: \wh{{\sf z}} \rightarrow {\sf z} = \cal{T}_{\epsilon}^{-1}\wh{{\sf z}}$. These transformations are expressed as asymptotic expansions in powers of a small dimensionless ordering parameter $\epsilon$ representing either a space-time scale ordering or the strength of the perturbation (e.g., wave amplitude) as \cite{RGL_82}
\begin{equation}
\left. \begin{array}{rcl}
\wh{z}^{a} & = & z^{a} \;+\; \epsilon\;G_{1}^{a} \;+\; \epsilon^{2} \left( G_{2}^{a} \;+\; \frac{1}{2}\,
{\sf G}_{1}\cdot\exd G_{1}^{a} \right) \;+\; \cdots \\
 &  & \\
z^{a} & = & \wh{z}^{a} \;-\; \epsilon\;G_{1}^{a} \;-\; \epsilon^{2} \left( G_{2}^{a} \;-\; 
\frac{1}{2}\,{\sf G}_{1}\cdot\exd G_{1}^{a} \right) \;+\; \cdots \\
\end{array} \right\}.
\label{eq:time_pst}
\end{equation}
Here, the $n$th-order vector field ${\sf G}_{n}\cdot\exd = G_{n}^{b}\partial_{b}$ is used to eliminate the fast space-time scales at order 
$\epsilon^{n}$. Note that in standard applications of Lie-transform perturbation theory, the time coordinate is unchanged (i.e., $\wh{t} \equiv t$) so 
that $G_{n}^{t} \equiv 0$ at all orders.

\subsection{Reduced Push-forward and Pull-back Operators}

The near-identity transformation (\ref{eq:time_pst}) induces a transformation of the extended Vlasov equation (\ref{eq:extVlasoveq_def}) as follows. First, the push-forward operator ${\sf T}_{\epsilon}^{-1}$ associated with the inverse phase-space transformation $\cal{T}_{\epsilon}^{-1}$, defined in Eq.~(\ref{eq:time_pst}), leads to the scalar-invariance relation
\begin{equation}
\wh{{\cal F}}(\wh{{\sf z}}) \;=\; \cal{F}({\sf z}) \;=\; \cal{F}\left(\cal{T}_{\epsilon}^{-1}\wh{{\sf z}}\right) \;\equiv\; 
{\sf T}_{\epsilon}^{-1}\cal{F}(\wh{{\sf z}}) 
\label{eq:scalar_relation}
\end{equation}
between the particle Vlasov distribution $\cal{F}$ on the extended particle phase space and the reduced Vlasov distribution $\wh{{\cal F}}$ on the extended reduced phase space. 

Next, we construct the reduced extended Vlasov operator $d_{\epsilon}/d\tau$, defined in terms of the push-forward operator ${\sf T}_{\epsilon}^{-1}$ (and its inverse, the pull-back operator ${\sf T}_{\epsilon}$) acting on an arbitrary function $G$ as 
\[ \frac{d_{\epsilon}G}{d\tau} \;\equiv\; {\sf T}_{\epsilon}^{-1}\left[\frac{d}{d\tau} ({\sf T}_{\epsilon}\,G)\right] \;\equiv\; \{ G,\; 
\wh{{\cal H}}\}_{\epsilon}. \]
In the last expression, $d_{\epsilon}/d\tau$ is expressed in terms of the extended reduced Hamiltonian $\wh{{\cal H}}$ and the extended reduced Poisson bracket
\begin{equation} 
\left\{ \wh{F},\; \wh{G}\right\}_{\epsilon} \;\equiv\; {\sf T}_{\epsilon}^{-1}\left(\left\{ {\sf T}_{\epsilon}\wh{F},\;{\sf T}_{\epsilon}\wh{G} \right\}\right).
\label{eq:reduced_PB_gen}
\end{equation}
The reduced Poisson bracket (\ref{eq:reduced_PB_gen}) can also be constructed from the reduced phase-space Lagrangian $\wh{\vb{\omega}}_{\epsilon} = 
\exd\wh{\Gamma}_{\epsilon} = \exd\left( {\sf T}_{\epsilon}^{-1}\Gamma + \exd\cal{S}\right) = {\sf T}_{\epsilon}^{-1}(\exd\Gamma) = 
{\sf T}_{\epsilon}^{-1}\vb{\omega}$, so that $\wh{J}_{\epsilon}^{ab} = (\wh{\vb{\omega}}_{\epsilon}^{\;\;-1})^{ab} \equiv \{ \wh{z}^{a},\; \wh{z}^{b}\}_{\epsilon}$, where we used the fact that $\exd^{2}{\cal S} \equiv 0$ (i.e., $\nabla\btimes\nabla{\cal S} \equiv 0$ in three dimensions) and the exterior derivative $\exd$ commutes with the operators ${\sf T}_{\epsilon}$ and ${\sf T}_{\epsilon}^{-1}$. 

Using the reduced extended Vlasov operator $d_{\epsilon}/d\tau$ induced by the near-identity phase-space transformation (\ref{eq:time_pst}), the extended reduced Hamilton's equations are expressed as
\begin{equation}
\frac{d_{\epsilon}\wh{z}^{a}}{d\tau} \;\equiv\; \left\{ \wh{z}^{a},\; \wh{{\cal H}} \right\}_{\epsilon},
\label{eq:redHameq}
\end{equation}
where the extended reduced Hamiltonian 
\begin{equation}
\wh{{\cal H}} \;\equiv\; {\sf T}_{\epsilon}^{-1}\cal{H} \;\equiv\; \wh{H} - \wh{w} 
\label{eq:redHam_def}
\end{equation}
is defined as the push-forward of the extended Hamiltonian $\cal{H} \equiv H - w$, with $\wh{H} \equiv {\sf T}_{\epsilon}^{-1}H - 
\partial\cal{S}/\partial t$. For many practical applications of reduced plasma dynamics, however, the new energy-momentum coordinates $\wh{p}_{\mu} = (-\,\wh{w}/c, \wh{p}_{\mu})$ are canonical coordinates and, therefore, the reduced Poisson bracket $\{\;,\;\}_{\epsilon} \equiv \{\;,\;\}_{\rm c}$ is the canonical bracket $\{\wh{F},\;\wh{G}\}_{\rm c} = (\partial \wh{F}/\partial\wh{x}^{\mu})\,
(\partial \wh{G}/\partial\wh{p}_{\mu}) - (\partial \wh{F}/\partial\wh{p}_{\mu})\,(\partial \wh{G}/\partial\wh{x}^{\mu})$.

Lastly, the near-identity phase-space transformation (\ref{eq:time_pst}) also introduces the reduced-displacement vector 
\begin{equation}
\vb{\rho}_{\epsilon} \;\equiv\; {\sf T}_{\epsilon}^{-1}{\bf x} \;-\; \wh{{\bf x}}, 
\label{eq:rho_epsilon}
\end{equation}
defined as the difference between the push-forward ${\sf T}_{\epsilon}^{-1}{\bf x}$ of the particle position ${\bf x}$ and the reduced position 
$\wh{{\bf x}}$. The reduced displacement (\ref{eq:rho_epsilon}) is expressed as
\begin{equation}
\vb{\rho}_{\epsilon} \;=\; -\,\epsilon\;G_{1}^{{\bf x}} \,-\, \epsilon^{2} \left( G_{2}^{{\bf x}} - \frac{1}{2}\,
{\sf G}_{1}\cdot\exd G_{1}^{{\bf x}} \right) + \cdots
\label{eq:rhoepsilon_def}
\end{equation}
in terms of the generating vector fields $({\sf G}_{1},{\sf G}_{2}, \cdots)$ associated with the near-identity phase-space transformation 
(\ref{eq:time_pst}). This reduced displacement plays in important role in the expression of polarization and magnetization effects in reduced plasma dynamics (see Sec.~\ref{subsec:pf_pond}).

\subsection{Reduced Vlasov-Maxwell Equations}

The extended reduced Vlasov equation for the extended reduced Vlasov distribution $\wh{{\cal F}}$ is expressed as
\begin{equation}
0 \;\equiv\; \left\{ \wh{{\cal F}},\; \wh{{\cal H}} \right\}_{\rm c} \;=\; \frac{d_{\epsilon}\wh{{\cal F}}}{d\tau} \;=\; 
\frac{d_{\epsilon}\wh{z}^{a}}{d\tau}\;\pd{\wh{{\cal F}}}{\wh{z}^{a}}.
\label{eq:redextVlasov_def}
\end{equation}
Here, the reduced Vlasov distribution is defined as the push-forward of the particle Vlasov distribution
\begin{equation}
\wh{{\cal F}}(\wh{{\sf z}}) \;\equiv\; c\,\delta[\wh{w} - \wh{H}(\wh{{\bf x}}, \wh{{\bf p}}, t)]\; \wh{F}(\wh{{\bf x}}, \wh{{\bf p}}, t),
\label{eq:ovF_extended} 
\end{equation}
with the reduced extended Hamiltonian (\ref{eq:redHam_def}) satisfying the physical constraint $\wh{{\cal H}} \equiv 0$.
When integrated over the reduced energy coordinate $\wh{w}$, Eq.~(\ref{eq:redextVlasov_def}) becomes the regular reduced Vlasov equation
\begin{equation}
0 \;=\; \frac{d_{\epsilon}\wh{F}}{dt} \;\equiv\; \pd{\wh{F}}{t} \;+\; \frac{d_{\epsilon}\wh{{\bf x}}}{dt}\bdot\pd{\wh{F}}{\wh{{\bf x}}} \;+\; \frac{d_{\epsilon}\wh{{\bf p}}}{dt}\bdot\pd{\wh{F}}{\wh{{\bf p}}},
\label{eq:oscVlasov_eq} 
\end{equation}
where we used $dt/d\tau = 1$ to eliminate $\tau$.

The dynamical reduction associated with the phase-space transformation (\ref{eq:time_pst}) introduces polarization and magnetization effects into the Maxwell equations, which transforms the microscopic Maxwell's equations (\ref{eq:E_Poisson})-(\ref{eq:B_Ampere}) into the macroscopic (reduced) Maxwell's equations 
\cite{Brizard_08,Ye_K}
\begin{eqnarray}
\nabla\bdot{\bf D}(x) & = & 4\pi\,\wh{\rho}(x), \label{eq:D_Poisson} \\
\nabla\btimes{\bf H}(x) \;-\; \frac{1}{c}\,\pd{{\bf D}}{t}(x) & = & \frac{4\pi}{c}\,\wh{{\bf J}}(x).
\label{eq:H_Ampere}
\end{eqnarray} 
Here, the terms ``microscopic'' and ``macroscopic'' are used to emphasize the fact that, while the sources $(\rho, {\bf J})$ of the microscopic Maxwell equations (\ref{eq:E_Poisson})-(\ref{eq:B_Ampere}) are expressed in terms of moments of the particle Vlasov distribution ${\cal F}$, the sources 
$(\wh{\rho}, \wh{{\bf J}})$ of the macroscopic Maxwell equations (\ref{eq:D_Poisson})-(\ref{eq:H_Ampere}) are expressed in terms of moments of the reduced Vlasov distribution $\wh{{\cal F}}$. In Eqs.~(\ref{eq:D_Poisson})-(\ref{eq:H_Ampere}), the microscopic electric and magnetic fields ${\bf E}$ and ${\bf B}$ are replaced by the macroscopic fields \cite{JDJ}
\begin{equation}
\left. \begin{array}{rcl}
{\bf D}(x) & = & {\bf E}(x) \;+\; 4\pi\,{\bf P}(x) \\
 &  & \\
{\bf H}(x) & = & {\bf B}(x) \;-\; 4\pi\,{\bf M}(x)
\end{array} \right\},
\label{eq:DH_def}
\end{equation}
where ${\bf P}$ and ${\bf M}$ are the polarization and magnetization associated with the dynamical reduction (\ref{eq:time_pst}). In the present paper, these polarization and magnetization are derived either by variational method (based on the existence of a reduced Lagrangian density) or by push-forward (Lie-transform) method (which establishes the connection between fluid moments in particle phase space and fluid moments in reduced phase space).

By comparing the microscopic and macroscopic (reduced) Maxwell's equations, we note that the dynamical reduction associated with the phase-space transformation (\ref{eq:time_pst}) has introduced the following expressions for the charge and current densities:
\begin{eqnarray}
\rho(x) & \equiv & \wh{\rho}(x) \;-\; \nabla\bdot{\bf P}(x), \label{eq:rho_Rpol} \\
{\bf J}(x) & \equiv & \wh{{\bf J}}(x) \;+\; \pd{{\bf P}}{t}(x) \;+\; c\,\nabla\btimes{\bf M}(x), \label{eq:J_Rpolmag}
\end{eqnarray}
where $\rho_{{\rm pol}} \equiv -\,\nabla\bdot{\bf P}$ denotes the polarization density, ${\bf J}_{{\rm pol}} \equiv \partial{\bf P}/\partial t$ denotes the polarization current, and ${\bf J}_{{\rm mag}} \equiv c\,\nabla\btimes{\bf M}$ denotes the magnetization current. Note that the reduced charge-current densities $(\wh{\rho}, \wh{{\bf J}})$ satisfy the reduced charge conservation law
\begin{eqnarray}
\pd{\wh{\rho}}{t} \;+\; \nabla\bdot\wh{{\bf J}} & = & \pd{\rho}{t} \;+\; \pd{}{t}\left(\nabla\bdot{\bf P}\frac{}{}\right) \;+\; \nabla\bdot{\bf J} \;-\; \nabla\bdot\left( \pd{{\bf P}}{t} \;+\; c\;\nabla\btimes{\bf M} \right) \nonumber \\
 & = & \pd{\rho}{t} \;+\; \nabla\bdot{\bf J} \;\equiv\; 0.
\label{eq:charge_red}
\end{eqnarray}
The macroscopic (reduced) Maxwell's equations (\ref{eq:D_Poisson})-(\ref{eq:H_Ampere}) can also be written in terms of the microscopic fields $({\bf E},
{\bf B})$ as
\begin{eqnarray}
\nabla\bdot{\bf E} & = & 4\pi\,\bigl( \wh{\rho} \;-\; \nabla\bdot{\bf P} \bigr), \label{eq:ED_Poisson} \\
\nabla\btimes{\bf B} \;-\; \frac{1}{c}\,\pd{{\bf E}}{t} & = & \frac{4\pi}{c}\,\left( \wh{{\bf J}} \;+\; \pd{{\bf P}}{t} \;+\; c\,\nabla\btimes{\bf M}
\right),
\label{eq:BH_Ampere}
\end{eqnarray}
where the polarization charge density and polarization-magnetization currents appear explicitly. It is this microscopic form that is most often useful for practical applications of reduced plasma dynamics \cite{Brizard_NFLR}.

\subsection{Reduced Vlasov-Maxwell Variational Principle}

We now show that the reduced Vlasov-Maxwell equations (\ref{eq:redextVlasov_def}) and (\ref{eq:ED_Poisson})-(\ref{eq:BH_Ampere}) can be derived from the reduced variational principle $\int d^{4}x\;\delta\wh{\cal{L}} = 0$, where the reduced Lagrangian density is
\begin{equation}
\wh{\cal{L}}(x) \;\equiv\; \frac{1}{16\pi}\,{\sf F}(x):{\sf F}(x) \;-\; \sum\;\int d^{4}\wh{p}\;\wh{{\cal F}}(x,\wh{p})\;\wh{{\cal H}}(x,
\wh{p}; A, {\sf F}).
\label{eq:redLag_def}
\end{equation}
Note that, as a result of the dynamical reduction of the Vlasov equation, the reduced Hamiltonian appearing in Eq.~(\ref{eq:redLag_def}) is expressed in terms of canonical energy-momentum coordinates as
\begin{equation}
\wh{{\cal H}}(x,\wh{p}; A, {\sf F}) \;\equiv\; \left( \frac{1}{2m}\,|\wh{{\bf p}} - \frac{e}{c}\,{\bf A}(x)|^{2} + e\,\Phi(x) - \wh{w}\right) \;+\; 
\Psi_{\epsilon}\left(\wh{{\bf p}} - \frac{e}{c}\,{\bf A}(x); {\sf F}(x)\right),
\label{eq:ovH_def}
\end{equation}
where the reduced {\it ponderomotive} potential $\Psi_{\epsilon}$ depends explicitly on the field tensor ${\sf F}_{\mu\nu}(x)$. From these field dependences, we define the reduced four-current density
\begin{equation}
\wh{J}^{\mu} \;=\; (c\wh{\rho}, \wh{{\bf J}}) \;=\; -\,c\,\;\sum\;\int d^{4}\wh{p}\;\wh{{\cal F}}\;\pd{\wh{{\cal H}}}{A_{\mu}} \;\equiv\;
\sum\;e\;\int d^{4}\wh{p}\;\wh{{\cal F}}\;\frac{d_{\epsilon}\wh{x}^{\mu}}{dt},
\label{eq:redJ_var}
\end{equation}
where 
\begin{equation}
\frac{d_{\epsilon}\wh{x}^{\mu}}{dt} \;\equiv\; \pd{\wh{{\cal H}}}{\wh{p}_{\mu}} \;=\; \left(c,\, \frac{1}{m}\,(\wh{{\bf p}} - \frac{e}{c}\,{\bf A}) + 
\pd{\Psi_{\epsilon}}{\wh{{\bf p}}} \right).
\label{eq:dx_epsilon}
\end{equation}
The reduced antisymmetric polarization-magnetization tensor \cite{BMB}, on the other hand, is defined as
\begin{equation}
{\sf K}^{\mu\nu} \;\equiv\; -\;2\;\sum\;\int d^{4}\wh{p}\;\wh{{\cal F}}\;\pd{\Psi_{\epsilon}}{{\sf F}_{\mu\nu}},
\label{eq:redH_var}
\end{equation}
where the polarization and magnetization ${\sf K}^{0i} = P^{i}$ and ${\sf K}^{ij} = \epsilon^{ijk}\,M_{k}$ are defined as  
\begin{equation}
\left( {\bf P},\; {\bf M} \right) \;\equiv\; -\;\sum\;\int d^{4}\wh{p}\;\wh{{\cal F}}\;
\left( \pd{\Psi_{\epsilon}}{{\bf E}},\; \pd{\Psi_{\epsilon}}{{\bf B}} \right). 
\label{eq:PM_var}
\end{equation}

\subsubsection{Reduced Vlasov-Maxwell Variational Principle.}

By following the same steps outlined in Sec.~\ref{subsec:VP_exact}, we begin our variational formulation of the reduced Vlasov-Maxwell equations with an expression for the variation of the Lagrangian density
\begin{equation}
\delta\wh{\cal{L}} \;=\; \partial_{\mu}\wh{\Lambda}^{\mu} \;-\; \sum\;\int d^{4}\wh{p}\;\wh{\cal{S}}\;\bigl\{ \wh{{\cal F}},\; \wh{{\cal H}} \bigr\} \;+\; \frac{\delta A_{\nu}}{4\pi} \left[\; \partial_{\mu}\bigl( {\sf F}^{\mu\nu} \;+\; 4\pi\, {\sf K}^{\mu\nu} \bigr) \;+\; \frac{4\pi}{c}\;\wh{J}^{\nu} \;\right], 
\label{eq:delLag_def} 
\end{equation}
where the reduced Noether four-vector 
\begin{equation}
\wh{\Lambda}^{\mu} \;\equiv\; \sum\;\int d^{4}\wh{p}\; \wh{\cal{S}}\,\wh{{\cal F}}\;\pd{\wh{{\cal H}}}{\wh{p}_{\mu}} \;+\; \frac{\delta A_{\nu}}{4\pi} 
\bigl( {\sf F}^{\nu\mu} + 4\pi\, {\sf K}^{\nu\mu} \bigr)
\label{eq:Noether_def}
\end{equation}
does not contribute to the variational principle $\int d^{4}x\,\delta\wh{\cal{L}} = 0$. As a result of the reduced variational principle, where the variations $\wh{\cal{S}}$ and $\delta A_{\nu}$ are arbitrary, we obtain the reduced Vlasov equation 
(\ref{eq:redextVlasov_def}) and the reduced ({\it macroscopic}) Maxwell equations
\begin{equation}
\pd{}{x^{\mu}} \left( {\sf F}^{\mu\nu} \;+\; 4\pi\,{\sf K}^{\mu\nu} \right) \;=\; -\;\frac{4\pi}{c}\;\wh{J}^{\nu},
\label{eq:Maxwell_var}
\end{equation}
from which we recover the reduced Maxwell equations (\ref{eq:ED_Poisson}) and (\ref{eq:BH_Ampere}). In addition, we find that the reduced charge conservation law (\ref{eq:charge_red}) follows immediately from the reduced Maxwell equations (\ref{eq:Maxwell_var}), since the tensors
${\sf F}^{\mu\nu}$ and ${\sf K}^{\mu\nu}$ are antisymmetric.

\subsubsection{Reduced Energy-Momentum Conservation Law.}

We now derive the energy-momentum conservation law from the reduced Noether equation $\delta\wh{\cal{L}} \equiv \partial_{\mu}\wh{\Lambda}^{\mu}$ associated with space-time translations $x^{\mu} \rightarrow x^{\mu} + \delta x^{\mu}$. The corresponding Eulerian variations 
$(\wh{\cal{S}}, \delta A_{\nu}, \delta\wh{{\cal L}})$ are 
\begin{equation}
\left. \begin{array}{rcl}
\wh{{\cal S}} & \equiv & \wh{p}_{\mu}\,\delta x^{\mu}  \\
\delta A_{\mu} & \equiv & {\sf F}_{\mu\nu}\;\delta x^{\nu} - \partial_{\mu}(A_{\nu}\,\delta x^{\nu}) \\
\delta\wh{{\cal L}} & \equiv & -\;\partial_{\mu}( \delta x^{\mu}\;\wh{{\cal L}})
\end{array} \right\}.
\label{eq:variations_red}
\end{equation}
After some cancellations introduced through the reduced Maxwell equations (\ref{eq:Maxwell_var}), we obtain the reduced energy-momentum conservation law 
$\partial_{\mu}{\sf T}^{\mu\nu} \equiv 0$, where the reduced stress-energy tensor is defined as
\begin{equation}
{\sf T}^{\mu\nu} \;\equiv\; \frac{1}{4\pi} \left[\; \frac{g^{\mu\nu}}{4}\; {\sf F}:{\sf F} \;-\; \left( {\sf F}^{\mu\sigma} + 4\pi\,
{\sf K}^{\mu\sigma}\right)\;{\sf F}_{\sigma}^{\;\;\nu} \;\right] \;+\; \sum\;\int d^{4}\wh{p}\; \pd{\wh{{\cal H}}}{\wh{p}_{\mu}}\;
\left(\wh{p}^{\nu} - \frac{e}{c}\,A^{\nu} \right)\;\wh{{\cal F}} 
\label{eq:Tmunu_def}
\end{equation}
naturally includes the polarization and magnetization (\ref{eq:PM_var}). An explicit proof of the reduced energy-momentum conservation law is given in Ref.~\cite{Brizard_08}. Note that, because of polarization and magnetization effects $(\Psi_{\epsilon} \neq 0)$, the reduced stress-energy tensor 
(\ref{eq:Tmunu_def}) does not appear to be symmetric (while it may be so when evaluated explicitly). The physical meaning of this unsymmetrical Minkowski form \cite{JDJ} for the reduced stress-energy tensor has a rich history in physics (briefly discussed by Boghosian \cite{BMB} and recently reviewed in Refs.~\cite{T_1,T_2}) but its further discussion is beyond the scope of this paper.

\subsection{\label{subsec:pf_pond}Push-forward Definitions of Polarization and Magnetization}

A complementary method for deriving the polarization and magnetization (\ref{eq:PM_var}) is provided by the push-forward method
\cite{Brizard_08}. From the scalar-invariance relation (\ref{eq:scalar_relation}), we construct the push-forward relation of fluid moments of an arbitrary function $\chi$ on particle phase space as follows. First, we transform the momentum average $[\chi]$ of an arbitrary phase-space function 
$\chi$:
\begin{eqnarray}
n\;[\chi] & \equiv & \int d^{3}p\;f\,\chi \;=\; \int d^{8}z\;\cal{F}\,\delta^{4}(x - r)\,\chi \nonumber \\
 & = & \int d^{8}\wh{z}\;\wh{{\cal F}}\;\;{\sf T}_{\epsilon}^{-1}\left[\delta^{4}(x - r)\;\chi \right] \nonumber \\
 & \equiv & \int d^{6}\wh{z}\;\wh{F}\; \delta^{3}(\wh{{\bf x}} + \vb{\rho}_{\epsilon} - {\bf r})\,{\sf T}_{\epsilon}^{-1}\chi,
\label{eq:chi_def}
\end{eqnarray}
where the extended Vlasov distribution (\ref{eq:ovF_extended}) was used ($n$ denotes the particle fluid density) and energy-time integrations were carried out. The reduced displacement (\ref{eq:rhoepsilon_def}) has components that are dependent on the fast space-time scales and components that are independent. Next, by expanding the right side of Eq.~(\ref{eq:chi_def}) in powers of $\vb{\rho}_{\epsilon}$ and integrating by parts, we obtain the push-forward relation
\begin{equation} 
n\;[\chi] \;\equiv\; \wh{n}\;\left[{\sf T}_{\epsilon}^{-1}\chi\right]^{\wedge} \;-\; \nabla\bdot\left( \wh{n}\;\left[\vb{\rho}_{\epsilon}\;
{\sf T}_{\epsilon}^{-1}\chi\right]^{\wedge} \;-\; \nabla\bdot\;\frac{\wh{n}}{2}\;\left[ \vb{\rho}_{\epsilon}\vb{\rho}_{\epsilon}\;{\sf T}_{\epsilon}^{-1}\chi\right]^{\wedge} \;+\; \cdots \right),
\label{eq:chi_push}
\end{equation}
where $[\cdots]^{\wedge}$ denotes an average with respect to the reduced Vlasov distribution $\wh{F}$ in reduced phase space ($\wh{n}$ denotes the reduced fluid density). In Eq.~(\ref{eq:chi_push}), the terms in the divergence on the right side include the dipolar contribution (linear in 
$\vb{\rho}_{\epsilon}$) and the quadrupolar contribution (quadratic in $\vb{\rho}_{\epsilon}$). We now apply the push-forward relation 
(\ref{eq:chi_push}) to derive dipolar expressions for the polarization and magnetization (where the quadrupolar and higher-order multipole contributions are ignored). 

First, we consider the case $\chi = e\;\;(= {\sf T}_{\epsilon}^{-1}\chi)$ so that the push-forward relation (\ref{eq:chi_push}) yields the expression for the charge density
\begin{equation} 
\rho \;=\; \wh{\rho} \;-\; \nabla\bdot\left( \sum\,e\, \wh{n}\;[\vb{\rho}_{\epsilon}]^{\wedge} \;+\; \cdots \right),
\label{eq:rho_push}
\end{equation}
where $\wh{\rho} \equiv \sum\,e\,\wh{n}$ defines the reduced charge density and higher-order multipole terms (e.g., quadrupole term that is quadratic in 
$\vb{\rho}_{\epsilon}$) are not displayed. By comparing Eq.~(\ref{eq:rho_Rpol}) with the relation (\ref{eq:rho_push}) between the particle charge density $\rho$ and the reduced charge density $\wh{\rho}$, we find the expression for the polarization (dipole-moment density)
\begin{equation}
{\bf P} \;\equiv\; \left. \left. \sum\, \right( \wh{n}\; [\vb{\pi}_{\epsilon}]^{\wedge} \right) \;=\; \sum\;\int d^{4}\wh{p}\; \vb{\pi}_{\epsilon}\;
\wh{{\cal F}},
\label{eq:Pol_push}
\end{equation}
where the reduced electric-dipole moment 
\begin{equation}
\vb{\pi}_{\epsilon} \;\equiv\; e\;\vb{\rho}_{\epsilon}
\label{eq:pi_def}
\end{equation}
is associated with the charge displacement induced by the phase-space transformation (\ref{eq:time_pst}). Note that the fast-time average $\langle
\vb{\pi}_{\epsilon}\rangle = e\,\langle\vb{\rho}_{\epsilon}\rangle$ of the reduced electric-dipole moment (\ref{eq:pi_def}) may not vanish and thus may provide important polarization effects \cite{Kaufman_gc,Brizard_NFLR}. 

Next, we consider the case $\chi = e{\bf v} \equiv e\,d{\bf x}/dt$, where the Lagrangian representation for the particle's velocity ${\bf v} = d{\bf x}
/dt$ is used. To obtain an expression for the push-forward ${\sf T}_{\epsilon}^{-1}(e\,d{\bf x}/dt)$, we use the reduced Vlasov operator $d_{\epsilon}
/dt$, defined in Eq.~(\ref{eq:redextVlasov_def}), to obtain
\begin{equation} 
e\,{\sf T}_{\epsilon}^{-1}{\bf v} \;=\; e\,{\sf T}_{\epsilon}^{-1}\,\frac{d{\bf x}}{dt} \;=\; e \left( 
{\sf T}_{\epsilon}^{-1}\,\frac{d}{dt}\, {\sf T}_{\epsilon} \right) {\sf T}_{\epsilon}^{-1}{\bf x} \;\equiv\; e\,\frac{d_{\epsilon}}{dt}\left({\sf T}_{\epsilon}^{-1}{\bf x}\right) \;=\; e\,\left( 
\frac{d_{\epsilon}\wh{{\bf x}}}{dt} \;+\; \frac{d_{\epsilon}\vb{\rho}_{\epsilon}}{dt} \right),
\label{eq:push_v}
\end{equation}
where we used Eq.~(\ref{eq:rho_epsilon}) to obtain the last expression. The reduced (e.g., guiding-center) velocity $d_{\epsilon}\wh{{\bf x}}/dt$ is independent of the fast space-time scales, while the reduced displacement velocity $d_{\epsilon}\vb{\rho}_{\epsilon}/dt$ exhibits both fast and slow space-time scales. Note that the fast-time-average particle {\it polarization} velocity $d_{\epsilon}\langle\vb{\rho}_{\epsilon}\rangle/dt$ may not vanish \cite{Sosenko} and thus may represent additional reduced dynamical effects (e.g., the standard polarization drift in guiding-center theory 
\cite{CB_08}) not included in the reduced velocity $d_{\epsilon}\wh{{\bf x}}/dt$. 

We now replace the reduced moment $[d_{\epsilon}\vb{\rho}_{\epsilon}/dt]^{\wedge}$ in the first term of Eq.~(\ref{eq:chi_push}) by taking the time derivative of the polarization (\ref{eq:Pol_push}) and using the reduced Vlasov equation (\ref{eq:oscVlasov_eq}) to obtain \cite{Brizard_08}
\begin{eqnarray} 
\pd{{\bf P}}{t} & = & \sum\,e\;\int d^{4}\wh{p} \left( \pd{\vb{\rho}_{\epsilon}}{t}\;\wh{{\cal F}} \;+\; \vb{\rho}_{\epsilon}\;\pd{\wh{{\cal F}}}{t} 
\right) \nonumber \\
 & = & \sum\,e\,\wh{n}\;\left[\,\frac{d_{\epsilon}\vb{\rho}_{\epsilon}}{dt}\right]^{\wedge} \;-\; \nabla\bdot\left( \sum\,e\,\wh{n}\;
\left[\frac{d_{\epsilon}\wh{{\bf x}}}{dt}\,\vb{\rho}_{\epsilon}\right]^{\wedge} \right),
\label{eq:partialP_partial t}
\end{eqnarray}
where integration by parts of the reduced Vlasov equation
\[ \pd{\wh{{\cal F}}}{t} \;=\; -\,\wh{\nabla}\wh{{\cal F}}\bdot \frac{d_{\epsilon}\wh{{\bf x}}}{dt} \;-\; \pd{\wh{{\cal F}}}{\wh{{\bf p}}}\bdot 
\frac{d_{\epsilon}\wh{{\bf p}}}{dt} \;\equiv\; -\;\wh{\nabla}\bdot\left( \wh{{\cal F}}\;\frac{d_{\epsilon}\wh{{\bf x}}}{dt}\right) \;-\;
\pd{}{\wh{{\bf p}}}\bdot\left( \wh{{\cal F}}\;\frac{d_{\epsilon}\wh{{\bf p}}}{dt} \right) \]
was carried out to obtain the last expression. Hence, the push-forward relation (\ref{eq:chi_push}) for $\chi = e\,d{\bf x}/dt$ yields the expression for the current density \cite{Brizard_08}
\begin{eqnarray}
{\bf J} & = & \wh{{\bf J}} \;+\; \pd{{\bf P}}{t} \;+\; \nabla\btimes\left( \sum\,e\,\wh{n}\;\left[\vb{\rho}_{\epsilon}\btimes
\frac{d_{\epsilon}\wh{{\bf x}}}{dt}\right]^{\wedge}\right) \;+\; \nabla\times\left( \sum\,\frac{e}{2}\,\wh{n}\;\left[\,\vb{\rho}_{\epsilon}\btimes\frac{d_{\epsilon}\vb{\rho}_{\epsilon}}{dt}\,\right]^{\wedge} \right) \nonumber \\
 & \equiv & \wh{{\bf J}} \;+\; \pd{{\bf P}}{t} \;+\; c\;\nabla\btimes{\bf M}
\label{eq:J_push}
\end{eqnarray}
where $\wh{{\bf J}} \equiv \sum\,e\,\wh{n}\;[d_{\epsilon}\wh{{\bf x}}/dt]^{\wedge}$ denotes the reduced current density, the second term represents the polarization current ${\bf J}_{{\rm pol}} \equiv \partial{\bf P}/\partial t$, and the third term represents the magnetization current ${\bf J}_{{\rm mag}} \equiv c\,\nabla\btimes{\bf M}$. By comparing Eq.~(\ref{eq:J_Rpolmag}) with the relation (\ref{eq:J_push}), we obtain the magnetization 
\begin{equation}
{\bf M} \;=\; \sum\;\frac{e}{c}\;\int d^{4}\wh{p}\; \vb{\rho}_{\epsilon}\btimes\left( \frac{1}{2}\;\frac{d_{\epsilon}\vb{\rho}_{\epsilon}}{dt} \;+\; \frac{d_{\epsilon}\wh{{\bf x}}}{dt} \right) \wh{{\cal F}} \;\equiv\; \sum\;\int d^{4}\wh{p}\; \left( \vb{\mu}_{\epsilon} \;+\; 
\frac{\vb{\pi}_{\epsilon}}{c}\btimes \frac{d_{\epsilon}\wh{{\bf x}}}{dt} \right) \wh{{\cal F}},
\label{eq:Mepsilon_def}
\end{equation}
which is represented as the sum (for each particle species) of an intrinsic magnetic-dipole contribution
\begin{equation}
\vb{\mu}_{\epsilon} \;\equiv\; \frac{e}{2c}\;\left(\vb{\rho}_{\epsilon}\btimes \frac{d_{\epsilon}\vb{\rho}_{\epsilon}}{dt}\right),
\label{eq:mu_def}
\end{equation}
and a moving electric-dipole \cite{JDJ} contribution $(\vb{\pi}_{\epsilon}\btimes d_{\epsilon}\wh{{\bf x}}/dt)$. We note that, since the relation 
$n^{\prime} \equiv \gamma\;n$ between the laboratory density $n^{\prime}$ and the rest-frame density $n$ involves the relativistic factor $\gamma =
1/\sqrt{1 - |{\bf v}|^{2}/c^{2}}$, the standard moving-dipole magnetization contribution \cite{pol_rel} to the polarization appearing in the charge density (\ref{eq:rho_push}) comes from the last term in the relativistic $c^{-2}$ correction
\begin{eqnarray}
e\,\vb{\rho}_{\epsilon}\;{\sf T}_{\epsilon}^{-1}(\gamma - 1) & = & \frac{e}{2c^{2}}\,\vb{\rho}_{\epsilon}\;\left|{\sf T}_{\epsilon}^{-1}{\bf v}
\right|^{2} \;+\; \cdots \nonumber \\
 & = & \frac{e}{2c^{2}}\;\vb{\rho}_{\epsilon} \left( \left|\frac{d_{\epsilon}\wh{{\bf x}}}{dt}\right|^{2} + \left|\frac{d_{\epsilon}\vb{\rho}_{\epsilon}}{dt}\right|^{2}\right) \;+\; \frac{e}{c^{2}}\;\frac{d_{\epsilon}\wh{{\bf x}}}{dt}\bdot\left(\frac{d_{\epsilon}\vb{\rho}_{\epsilon}}{dt}\;\vb{\rho}_{\epsilon}\right),
\label{eq:pol_rel}
\end{eqnarray}
where the first term yields a small relativistic correction to the electric dipole moment. We return to this relativistic magnetization effect in 
Sec.~\ref{subsec:pond_push} [see Eq.~(\ref{eq:P2_mu})].

A simple example of the magnetization (\ref{eq:mu_def}) is provided by guiding-center dynamics \cite{Kaufman_gc,CB_08}, where $\vb{\rho}$ denotes the gyroangle-dependent gyroradius vector (with $d_{\epsilon}\vb{\rho}_{\epsilon}/dt = \Omega\,\partial\vb{\rho}/\partial\theta$ to lowest order, where $\Omega$ denotes the gyrofrequency) and the guiding-center intrinsic magnetization $\vb{\mu}_{\rm gc} \equiv -\,\mu\,\bhat$ is defined in terms of the magnetic moment $\mu \equiv m|{\bf v}_{\bot}|^{2}/2B$ (an adiabatic invariant for guiding-center motion) and $\bhat \equiv {\bf B}/B$. The guiding-center electric dipole moment \cite{Kaufman_gc}, on the other hand, is expressed as $\vb{\pi}_{\rm gc} \equiv e\,(\bhat/\Omega)\btimes{\bf v}_{\rm gc}$, where 
${\bf v}_{\rm gc}$ denotes the guiding-center drift velocity (excluding the $E\times B$ velocity).

\subsection{Polarization and Magnetization}

We now summarize the main results of this Section. The polarization and magnetization 
\begin{eqnarray*} 
{\bf P} & = & -\;\sum\;\wh{n}\;\pd{[\Psi_{\epsilon}]^{\wedge}}{{\bf E}} \;=\; \sum\; \wh{n}\;[\vb{\pi}_{\epsilon}]^{\wedge}, \\
{\bf M} & = & -\;\sum\;\wh{n}\;\pd{[\Psi_{\epsilon}]^{\wedge}}{{\bf B}} \;=\; \sum\; \wh{n}\;\left[\vb{\mu}_{\epsilon} \;+\;
\frac{\vb{\pi}_{\epsilon}}{c}\btimes\frac{d_{\epsilon}\wh{{\bf x}}}{dt} \right]^{\wedge},
\end{eqnarray*}
can either be derived by variational method from the reduced potential $\Psi_{\epsilon}$ or by push-forward method from the reduced displacement $\vb{\rho}_{\epsilon} \equiv {\sf T}_{\epsilon}^{-1}{\bf x} - \wh{{\bf x}}$. By combining these expressions, we obtain the 
relations between the reduced dipole moments $(\vb{\pi}_{\epsilon}, \vb{\mu}_{\epsilon})$ and the partial derivatives $(\partial\Psi_{\epsilon}/
\partial{\bf E}, \partial\Psi_{\epsilon}/\partial{\bf B})$ of the reduced Hamiltonian:
\begin{equation}
\left. \begin{array}{rcl}
-\;\partial\Psi_{\epsilon}/\partial{\bf E} & \equiv & \vb{\pi}_{\epsilon} \\
 &  & \\
-\;\partial\Psi_{\epsilon}/\partial{\bf B} & \equiv & \vb{\mu}_{\epsilon} + (\vb{\pi}_{\epsilon}/c)\btimes d_{\epsilon}\wh{{\bf x}}/dt
\end{array} \right\}.
\label{eq:pimu_partial}
\end{equation}
These relations emphasize the complementarity of the reduced variational and push-forward methods.

\section{\label{sec:Lie}Oscillation-center dynamics in weakly-magnetized plasmas}

The problem of the low-frequency oscillation-center Hamiltonian dynamics of charged particles in high-frequency, short-wavelength electromagnetic waves is the paradigm for applications of Lie-transform perturbation methods in plasma physics \cite{GKL,Cary_Kaufman,HW,Kaufman_Holm}. Here, the fast wave space-time scales are removed asymptotically from the Hamiltonian dynamics by a time-dependent phase-space noncanonical transformation $z^{a} = (ct, {\bf x}; w/c, {\bf p}) \rightarrow \ov{z}^{a} = (ct, \ov{{\bf x}}; \ov{w}/c, \ov{{\bf p}})$, where $\ov{{\bf x}}$ denotes the oscillation-center position, and $(\ov{w},\, \ov{{\bf p}} \equiv m\,\ov{{\bf v}})$ denote the oscillation-center's kinetic energy-momentum. Note that, although the phase-space transformation is time-dependent, the time coordinate $t$ remains unchanged by the transformation (i.e., the particle and oscillation-center times are identical).

First, we decompose the electromagnetic potentials $(\Phi,{\bf A})$ in powers of the wave amplitude (represented by the ordering parameter $\epsilon$):
\begin{equation} 
\left( \begin{array}{c}
\Phi \\
{\bf A}
\end{array} \right) \;=\; \sum_{n = 0}^{\infty}\; \epsilon^{n}\; \left( \begin{array}{c}
\Phi_{n} \\
{\bf A}_{n}
\end{array} \right),
\label{eq:phiA_n}
\end{equation}
where the lowest-order background plasma is represented by the zeroth-order fields $(\Phi_{0},{\bf A}_{0})$, while the primary wave fields 
$({\bf E}_{1}, {\bf B}_{1})$ are represented by the first-order potentials $(\Phi_{1},{\bf A}_{1})$ and the second-order (back-reaction) electromagnetic fields $({\bf E}_{2},{\bf B}_{2})$ are represented by the second-order potentials $(\Phi_{2},{\bf A}_{2})$. These second-order fields, which represents the plasma response to the external first-order fields, contain contributions that are weakly space-time dependent compared to the rapid wave space-time scales, and contributions with rapid {\it second-harmonic} wave space-time scales.

The extended-phase-space Hamiltonian dynamics of charged particles in the electromagnetic fields (\ref{eq:phiA_n}) is expressed in terms of the extended Hamiltonian
\begin{equation}
\cal{H} \;=\; \frac{1}{2m}\,|{\bf p}|^{2} \;-\; w \;\equiv\; {\cal H}_{0},
\label{eq:Hpond_def}
\end{equation}
and the extended-phase-space Lagrangian
\begin{eqnarray}
\Gamma & = & \left[ {\bf p} + \frac{e}{c}\,\left( {\bf A}_{0} + \epsilon\,{\bf A}_{1} + \epsilon^{2}\,{\bf A}_{2} + \cdots \right)\right]
\bdot \exd{\bf x} \;-\; \left[ w + e \left( \Phi_{0} + \epsilon\,\Phi_{1} + \epsilon^{2}\,\Phi_{2} + \cdots \right)\right]\,\exd t \nonumber \\
 & \equiv & \Gamma_{0} \;+\; \epsilon\,\Gamma_{1} \;+\; \epsilon^{2}\,\Gamma_{2} \;+\; \cdots.
\label{eq:PSLpond_def}
\end{eqnarray}
Here, the unperturbed Poisson bracket $\{\;,\;\}_{0}$ is obtained from $\Gamma_{0}$ as
\begin{equation}
\{ F,\; G \}_{0} \;\equiv\; \left( \pd{F}{x^{\mu}}\;\pd{G}{p_{\mu}} \;-\; \pd{F}{p_{\mu}}\;\pd{G}{x^{\mu}} \right) + 
\frac{e}{c}\,{\sf F}_{(0)\mu\nu}\;\pd{F}{p_{\mu}}\;\pd{G}{p_{\nu}}.
\label{eq:extPB0_def}
\end{equation}

\subsection{Eikonal Representation}

We now introduce the eikonal representation for the electromagnetic fields (\ref{eq:phiA_n}). The previous work by Cary and Kaufman \cite{Cary_Kaufman} considered the case of an unmagnetized background plasma (with $\Phi_{0} = 0 = {\bf A}_{0}$) and the case of a strongly magnetized plasma (with $\Phi_{0} = 0$). Here, we consider the intermediate case of a weakly-magnetized background plasma described as follows.

First, we assume that the zeroth-order (background) fields $A^{\mu}_{(0)} \equiv (\Phi_{0},{\bf A}_{0})$ are weakly space-time dependent:
\begin{equation} 
A^{\mu}_{(0)} \;\equiv\; A^{\mu}_{(0)}(\epsilon_{0}x),
\label{eq:phiA_0}
\end{equation}
where $\epsilon_{0} \ll 1$ denotes the eikonal ordering parameter (warning: $\epsilon_{0}$ is not the permitivity of free space), so that the background electric and magnetic fields 
\begin{equation}
{\sf F}_{(0)\mu\nu} \;\equiv\; \epsilon_{0}\,{\sf F}_{(0)\mu\nu}(\epsilon_{0}x) 
\label{eq:EB_0}
\end{equation}
are ordered small; the weak-background-field (WBF) ordering (\ref{eq:EB_0}) is consistent with a weakly-magnetized quasi-static background plasma. In particular, according to this ordering, the wave frequency $\omega$ is considered much larger than the gyrofrequency $\Omega_{0} = eB_{0}/mc$ of a charged particle moving in this weak magnetic field while the wavelength $\lambda \equiv 2\pi\,|{\bf k}|^{-1}$ is considered much smaller than the particle's gyroradius $\rho_{0}$. Because of the mass dependence associated with this ordering, the weak-background-field (WBF) ordering (\ref{eq:EB_0}) is especially appropriate for ions (because of their large masses), while the ordering might be invalid for electrons (because of their small masses).

Next, the first-order wave fields are decomposed in terms of the eikonal representation:
\begin{equation}
\left( \begin{array}{c}
\Phi_{1} \\
{\bf A}_{1}
\end{array} \right) \;\equiv\; \left( \begin{array}{c}
\wt{\Phi}_{1} \\
\wt{{\bf A}}_{1}
\end{array} \right) e^{i\,\Theta/\epsilon_{0}} \;+\; \left( \begin{array}{c}
\wt{\Phi}_{1}^{*} \\
\wt{{\bf A}}_{1}^{*}
\end{array} \right) e^{-\,i\,\Theta/\epsilon_{0}},
\label{eq:first_eikonal}
\end{equation}
where the wave eikonal-amplitudes (denoted by a tilde) are weakly-varying space-time functions and derivatives of the eikonal phase 
$\Theta(\epsilon_{0}{\bf r},\epsilon_{0} t)$:
\begin{equation}
\epsilon_{0}^{-1}\pd{\Theta}{x^{\mu}}(\epsilon_{0}x) \;\equiv\; k_{\mu}(\epsilon_{0}x) \;=\; (-\omega/c,\; {\bf k})
\label{eq:eikonal_kw}
\end{equation}
define the weakly-varying wavevector ${\bf k}$ and wave frequency $\omega$. Note that the first-order electromagnetic field has the eikonal-amplitude
\begin{equation}
\wt{{\sf F}}_{(1)\mu\nu} \;\equiv\; i \left( k_{\mu}\;\wt{A}_{(1)\nu} \;-\; k_{\nu}\;\wt{A}_{(1)\mu} \right).
\label{eq:EB1_def}
\end{equation}

Lastly, the eikonal representation of the second-order fields is
\begin{equation}
\left( \begin{array}{c}
\Phi_{2} \\
{\bf A}_{2}
\end{array} \right) \;\equiv\; \left( \begin{array}{c}
\ov{\Phi}_{2} \\
\ov{{\bf A}}_{2}
\end{array} \right) \;+\; \left( \begin{array}{c}
\wt{\Phi}_{2} \\
\wt{{\bf A}}_{2}
\end{array} \right) e^{2i\,\Theta/\epsilon_{0}} \;+\; \left( \begin{array}{c}
\wt{\Phi}_{2}^{*} \\
\wt{{\bf A}}_{2}^{*}
\end{array} \right) e^{-\,2i\,\Theta/\epsilon_{0}},
\label{eq:second_eikonal}
\end{equation}
where the eikonal-averaged weakly-varying back-reaction fields $(\ov{\Phi}_{2},\ov{{\bf A}}_{2})$ are generated by the second-order 
\textit{ponderomotive} effects (as discussed below) while second-harmonic terms $(\wt{\Phi}_{2},\wt{{\bf A}}_{2})$ are, henceforth, ignored (i.e., an eikonal-averaged expression involving second-harmonic terms appears at the fourth-order in wave amplitude, which falls outside the scope of our work).

\subsection{Oscillation-center Transformation and Ponderomotive Hamiltonian}

As a result of the perturbation expansion (\ref{eq:phiA_n}) in the electromagnetic potentials $(\Phi, {\bf A})$, we see that the Hamiltonian 
(\ref{eq:Hpond_def}) exhibits explicit dependence on the eikonal phase $\Theta$ and, thus, the particle dynamics exhibits fast space-time scales in addition to the slow space-time scales associated with the background plasma.

Lie-transform Hamiltonian perturbation theory \cite{RGL_82} is concerned with the derivation of a new extended oscillation-center Hamiltonian
\begin{equation} 
\ov{{\cal H}} \;\equiv\; {\sf T}_{\epsilon}^{-1}{\cal H} \;=\; \left( \ov{H}_{0} \;+\; \epsilon\,\ov{H}_{1} \;+\; \epsilon^{2}\,\ov{H}_{2} \;+\; \cdots \right) \;-\; \ov{w},
\label{eq:Hamoc_def}
\end{equation}
where the first-order and second-order oscillation-center Hamiltonians are
\begin{eqnarray}
\ov{H}_{1} & = & H_{1} \;-\; {\sf G}_{1}\cdot\exd{\cal H}_{0} \;\equiv\; -\; {\sf G}_{1}\cdot\exd{\cal H}_{0}, \label{eq:Ham_1} \\
\ov{H}_{2} & = & H_{2} \;-\; {\sf G}_{2}\cdot\exd{\cal H}_{0} \;-\; \frac{1}{2}\;{\sf G}_{1}\cdot\exd( H_{1} + \ov{H}_{1}) \;\equiv\;
-\; {\sf G}_{2}\cdot\exd{\cal H}_{0} \;-\; \frac{1}{2}\;{\sf G}_{1}\cdot\exd\ov{H}_{1}.
\label{eq:Ham_2}
\end{eqnarray} 
The last expressions in Eqs.~(\ref{eq:Ham_1})-(\ref{eq:Ham_2}) use the fact that the Hamiltonian perturbation terms $H_{n}\;(n \geq 1)$ are identically zero in the Hamiltonian (\ref{eq:Hpond_def}). The generating vector fields $({\sf G}_{1}, {\sf G}_{2}, \cdots)$, on the other hand, are determined from the requirement that the oscillation-center phase-space Lagrangian
\begin{equation}
\ov{\Gamma} \;\equiv\; {\sf T}_{\epsilon}^{-1}\Gamma \;+\; \exd{\cal S} \;\equiv\; \ov{\Gamma}_{0},
\label{eq:ov_Gamma}
\end{equation}
retains its unperturbed form. From this requirement, we obtain
\begin{eqnarray}
G_{1}^{a} & = & \left\{ S_{1},\; z^{a}\right\}_{0} \;+\; \frac{e}{c}\,A_{(1)\mu}\;\left\{ x^{\mu},\; z^{a}\right\}_{0}, \label{eq:G1_def} \\
G_{2}^{a} & = & \left\{ S_{2},\; z^{a}\right\}_{0} \;+\; \frac{e}{c}\,A_{(2)\mu}\;\left\{ x^{\mu},\; z^{a}\right\}_{0} \;-\; \frac{e}{2c}\;
\left\{ S_{1},\; x^{\mu}\right\}_{0}\;{\sf F}_{(1)\mu\nu}\;\left\{ x^{\nu},\; z^{a}\right\}_{0},
\label{eq:G2_def}
\end{eqnarray}
where the scalar fields $(S_{1}, S_{2}, \cdots)$ are determined from Eqs.~(\ref{eq:Ham_1})-(\ref{eq:Ham_2}) by requiring that the oscillation-center Hamiltonian (\ref{eq:Hamoc_def}) be eikonal-phase independent.

\subsubsection{First-order analysis.}

From Eqs.~(\ref{eq:Ham_1}) and (\ref{eq:G1_def}), we obtain the first-order Hamiltonian
\begin{equation} 
\ov{H}_{1} \;\equiv\; -\;\left\{ S_{1},\; {\cal H}_{0}\right\}_{0} \;-\; \frac{e}{c}\,A_{(1)\mu}\;\left\{ x^{\mu},\; {\cal H}_{0}\right\}_{0} \;=\; 
-\,\frac{dS_{1}}{dt} \;+\; e \left( \Phi_{1} \;-\; \frac{{\bf v}}{c}\bdot{\bf A}_{1} \right) \;\equiv\; 0,
\label{eq:Ham1_def}
\end{equation}
where the first-order oscillation-center Hamiltonian $\ov{H}_{1}$ must vanish since $(\Phi_{1}, {\bf A}_{1}, S_{1})$ are all explicitly eikonal-phase dependent. By inserting the eikonal representation (\ref{eq:first_eikonal}) into Eq.~(\ref{eq:Ham1_def}), we obtain the eikonal equation for $\wt{S}_{1}$:
\begin{equation} 
-i\,\omega^{\prime}\,\wt{S}_{1} \;+\; \epsilon_{0}\,e\,\left( {\bf E}_{0} + \frac{{\bf v}}{c}\btimes{\bf B}_{0}\right)\bdot
\pd{\wt{S}_{1}}{{\bf p}} \;=\; e \left( \wt{\Phi}_{1} \;-\; \frac{{\bf v}}{c}\bdot\wt{{\bf A}}_{1} \right)
\label{eq:S1_eikonal}
\end{equation}
where $\omega^{\prime} = \omega - {\bf k}\bdot{\bf v}$ is the Doppler-shifted wave frequency, and the WBF eikonal ordering 
(\ref{eq:EB_0}) has been used. Next, by substituting the WBF-eikonal expansion 
\begin{equation}
\wt{S}_{1} \;=\; \wt{S}_{10} \;+\; \epsilon_{0}\,\wt{S}_{11} \;+\; \cdots
\label{eq:S_WBF}
\end{equation} 
into Eq.~(\ref{eq:S1_eikonal}), we obtain the lowest-order eikonal amplitude
\begin{equation}
\wt{S}_{10} \;=\; \frac{ie}{\omega^{\prime}} \left( \wt{\Phi}_{1} \;-\; \frac{{\bf v}}{c}\bdot
\wt{{\bf A}}_{1} \right), 
\label{eq:S10_def}
\end{equation}
and the first-order WBF-eikonal correction
\begin{equation}
\wt{S}_{11} \;=\; -\;\frac{ie}{\omega^{\prime}}\;\left( {\bf E}_{0} + \frac{{\bf v}}{c}\btimes{\bf B}_{0}\right)\bdot\wt{\vb{\xi}} \;\equiv\;
-\,i\;{\bf F}_{0}\bdot\frac{\wt{\vb{\xi}}}{\omega^{\prime}}, 
\label{eq:S11_def}
\end{equation}
where the first-order spatial displacement $\wt{\vb{\xi}}$ is defined in terms of the lowest-order eikonal amplitude 
\cite{HW}
\begin{equation} 
\wt{\vb{\xi}} \;\equiv\; \pd{\wt{S}_{10}}{{\bf p}} \;=\; -\;\frac{e}{m\omega^{\prime 2}} \;\left( \wt{{\bf E}}_{1} + \frac{{\bf v}}{c}\btimes
\wt{{\bf B}}_{1}\right).
\label{eq:xi_def}
\end{equation}
Note that the first-order spatial displacement $\wt{\vb{\xi}}$ also satisfies the eikonal relation $i\,m\omega^{\prime}\,\wt{\vb{\xi}} = i{\bf k}\,
\wt{S}_{10} + e\wt{{\bf A}}_{1}/c$, which can be expressed as
\begin{equation}
-\,m\;\frac{d_{0}\vb{\xi}}{dt} \;=\; \nabla S_{10} \;+\; \frac{e}{c}\;{\bf A}_{1},
\label{eq:dxi_dt}
\end{equation}
where $d_{0}/dt \equiv \partial/\partial t + {\bf v}\bdot\nabla$ and $\omega^{\prime} \equiv -\,\epsilon_{0}^{-1}d_{0}\Theta/dt$. 

The phase-space transformation (\ref{eq:time_pst}) can thus be expressed explicitly as
\begin{equation}
\left. \begin{array}{rcl}
\ov{{\bf x}} & = & {\bf x} - \epsilon\,\partial S_{10}/\partial{\bf p} + \cdots \;=\; {\bf x} \;-\; \epsilon\,\vb{\xi} \;+\; \cdots \\
 &  & \\
\ov{{\bf p}} & = & {\bf p} + \epsilon\,(\nabla S_{10} + e{\bf A}_{1}/c) + \cdots \;=\; {\bf p} \;-\; \epsilon\,m\,d_{0}\vb{\xi}/dt \;+\; \cdots \\
 &  & \\
\ov{w} & = & w - \epsilon\,(\partial S_{10}/\partial t - e\Phi_{1}) + \cdots \;=\; w - \epsilon\,m{\bf v}\bdot d_{0}\vb{\xi}/dt \;+\; \cdots
\end{array} \right\}.
\label{eq:xpw_osc}
\end{equation}
Hence, to first order in $\epsilon$, the oscillation-center momentum coordinate is 
\[ \ov{{\bf p}} \;=\; m\,\left({\bf v} - \epsilon\;\frac{d_{0}\vb{\xi}}{dt} \right) \;\equiv\; m\;\frac{d\ov{{\bf x}}}{dt}, \]
and the oscillation-center energy coordinate is 
\[ \ov{w} \;=\; \frac{m}{2}\,|{\bf v}|^{2} \;-\; \epsilon\;m{\bf v}\bdot \frac{d_{0}\vb{\xi}}{dt} \;\equiv\; \frac{1}{2m}\;|\ov{{\bf p}}|^{2}, \]
which satisfies the unperturbed constraint $\ov{{\cal H}}_{0} = |\ov{{\bf p}}|^{2}/2m - \ov{w} \equiv 0$ for the oscillation-center kinetic energy-momentum coordinates.

\subsubsection{Second-order analysis.}

Next, from Eqs.~(\ref{eq:Ham_2}), (\ref{eq:G2_def}), and (\ref{eq:Ham1_def}), the expression for the second-order Hamiltonian is
\begin{eqnarray} 
\ov{H}_{2} & \equiv & -\,\{ S_{2},\; {\cal H}_{0}\}_{0} \;-\; \frac{e}{c}\,A_{(2)\mu}\;\{x^{\mu},\; {\cal H}_{0}\}_{0} \;-\; \frac{e}{2c}\;
\pd{S_{1}}{p_{\mu}}\;{\sf F}_{(1)\mu\nu}\;\{ x^{\nu},\; {\cal H}_{0}\}_{0} \nonumber \\
 & = & -\;\frac{dS_{2}}{dt} \;+\; e \left( \Phi_{2} - \frac{{\bf v}}{c}\bdot{\bf A}_{2}\right) \;-\; \frac{e}{2}\;\left( {\bf E}_{1} + \frac{{\bf v}}{c}\btimes{\bf B}_{1}\right)\bdot\pd{S_{1}}{{\bf p}},
\label{eq:H2pond_def}
\end{eqnarray}
where the right side has both eikonal-dependent and eikonal-independent terms. To derive the eikonal-independent terms that define the second-order oscillation-center Hamiltonian $\ov{H}_{2}$, we perform an eikonal-phase average of the right side of Eq.~(\ref{eq:H2pond_def}) to obtain
\begin{equation}
\ov{H}_{2} \;=\; e \left( \ov{\Phi}_{2} \;-\; \frac{{\bf v}}{c}\bdot\ov{{\bf A}}_{2} \right) \;-\; \frac{e}{2}\;\left\langle\left( {\bf E}_{1} + 
\frac{{\bf v}}{c}\btimes{\bf B}_{1}\right)\bdot\pd{S_{1}}{{\bf p}} \right\rangle,
\label{eq:H2oc_def} 
\end{equation}
where the contributions of the second-harmonic eikonal-amplitudes $(\wt{\Phi}_{2}, \wt{{\bf A}}_{2}, \wt{S}_{2} \equiv S_{2})$ vanish. 

Using the definition (\ref{eq:xi_def}), the lowest-order contribution from the second term in Eq.~(\ref{eq:H2oc_def}) is
\begin{equation}
-\; \frac{e}{2}\;\left\langle\left( {\bf E}_{1} + \frac{{\bf v}}{c}\btimes{\bf B}_{1}\right)\bdot\pd{S_{10}}{{\bf p}} \right\rangle \;=\; 
m\omega^{\prime\,2}\;|\wt{\vb{\xi}}|^{2},
\label{eq:pond_20}
\end{equation}
which determines the standard ponderomotive Hamiltonian \cite{Cary_Kaufman}. Using the definition (\ref{eq:S11_def}), on the other hand, the first-order WBF correction is
\begin{eqnarray}
-\; \frac{e}{2}\;\left\langle\left( {\bf E}_{1} + \frac{{\bf v}}{c}\btimes{\bf B}_{1}\right)\bdot\pd{S_{11}}{{\bf p}} \right\rangle & = & 
\pd{}{{\bf p}}\bdot{\rm Re}\left( m\omega^{\prime\,2}\wt{\vb{\xi}}\;\wt{S}_{11}^{*}\right) \;=\; \pd{}{{\bf p}}\bdot\left[\; \frac{i}{2}\,m\omega^{\prime}\;{\bf F}_{0}\btimes\left(\wt{\vb{\xi}}\btimes\wt{\vb{\xi}}^{*}\right)\;\right] \nonumber \\
 & \equiv & -\;\left( {\bf E}_{0} + \frac{{\bf v}}{c}\btimes{\bf B}_{0}\right)\bdot\ov{\vb{\pi}}_{2} \;-\; {\bf B}_{0}\bdot\ov{\vb{\mu}}_{2},
\label{eq:pond_21}
\end{eqnarray}
where we introduced the definitions for the ponderomotive electric dipole moment
\begin{equation}
\ov{\vb{\pi}}_{2} \;\equiv\; e\;{\bf k}\btimes \left( i\,\wt{\vb{\xi}}\btimes\wt{\vb{\xi}}^{*} \right),
\label{eq:pi_2}
\end{equation}
and the ponderomotive magnetic dipole moment
\begin{equation} 
\ov{\vb{\mu}}_{2} \;\equiv\; \frac{e}{c}\;\omega^{\prime}\; \left( i\,\wt{\vb{\xi}}\btimes\wt{\vb{\xi}}^{*} \right).
\label{eq:mu_2}
\end{equation}
Note that the intrinsic ponderomotive magnetic-dipole moment $\ov{\vb{\mu}}_{2}$ is accompanied in Eq.~(\ref{eq:pond_21}) by the ponderomotive moving-electric-dipole moment $\ov{\vb{\pi}}_{2}\btimes{\bf v}/c$ \cite{JDJ}. As indicated above, however, the moving-magnetic-dipole contribution
$({\bf v}/c)\btimes\ov{\vb{\mu}}_{2}$ to the ponderomotive polarization appears as a relativistic correction, which is ignored in the present nonrelativistic treatment.

\subsubsection{Ponderomotive Hamiltonian.}

By combining Eqs.~(\ref{eq:pond_20})-(\ref{eq:pond_21}), the second-order oscillation-center Hamiltonian is defined by the eikonal-averaged terms on the right side of Eq.~(\ref{eq:H2pond_def}):
\begin{eqnarray}
\ov{H}_{2} & = & e \left( \ov{\Phi}_{2} \;-\; \frac{\ov{{\bf v}}}{c}\bdot\ov{{\bf A}}_{2} \right) \;+\; 
m\omega^{\prime 2}\;|\wt{\vb{\xi}}|^{2} \;-\; \epsilon_{0} \left[\; {\bf E}_{0}\bdot\ov{\vb{\pi}}_{2} \;+\; {\bf B}_{0}\bdot\left( \ov{\vb{\mu}}_{2} 
\;+\; \ov{\vb{\pi}}_{2}\btimes\frac{\ov{{\bf v}}}{c} \right) \;\right] \nonumber \\
 & \equiv & e \left( \ov{\Phi}_{2} \;-\; \frac{\ov{{\bf v}}}{c}\bdot\ov{{\bf A}}_{2} \right) \;+\; \ov{\Psi}_{2},
\label{eq:H2_oc}
\end{eqnarray}
where the reduced potential $\ov{\Psi}_{2} = \ov{\Psi}_{20} + \epsilon_{0}\,\ov{\Psi}_{21}$ contains the ponderomotive potential $\ov{\Psi}_{20}$ and its first-order (WBF) correction $\ov{\Psi}_{21}$, which includes the ponderomotive polarization and magnetization effects (\ref{eq:pi_2})-(\ref{eq:mu_2}). 

Lastly, by using the ponderomotive dipole moments (\ref{eq:pi_2}) and (\ref{eq:mu_2}), we can immediately write down expressions for the ponderomotive polarization and magnetization
\begin{equation}
\left( \ov{{\bf P}}_{2},\; \ov{{\bf M}}_{2} \right) \;\equiv\; \sum\; \int d^{3}\ov{p}\;\ov{F} \left( \ov{\vb{\pi}}_{2},\;
\ov{\vb{\mu}}_{2} \;+\; \ov{\vb{\pi}}_{2}\btimes\frac{\ov{{\bf v}}}{c} \right) \;=\; -\;\sum\; \int d^{3}\ov{p}\;\ov{F} \left( \pd{\ov{\Psi}_{21}}{{\bf E}_{0}},
\; \pd{\ov{\Psi}_{21}}{{\bf B}_{0}} \right),
\label{eq:PM2_push}
\end{equation}
which satisfy the relations (\ref{eq:pimu_partial}) between the variational and push-forward methods. Note that the polarization and magnetization 
\begin{equation} 
{\bf P} \;=\; \epsilon\,{\bf P}_{1} \;+\; \epsilon^{2}\,{\bf P}_{2} \;+\; \cdots \;\;\;{\rm and}\;\;\;
{\bf M} \;=\; \epsilon\,{\bf M}_{1} \;+\; \epsilon^{2}\,{\bf M}_{2} \;+\; \cdots 
\label{eq:PM_osc}
\end{equation}
have first-order contributions that vanish upon eikonal-phase averaging (i.e., $\langle {\bf P}_{1}\rangle = 0 = \langle {\bf M}_{1}\rangle$), while the eikonal-averaged contributions are second-order in wave amplitude: $\langle {\bf P}\rangle = \epsilon^{2}\,\ov{{\bf P}}_{2}$ and $\langle {\bf M}
\rangle = \epsilon^{2}\,\ov{{\bf M}}_{2}$.

\subsection{Oscillation-center Pull-back Operator}

The relation between the particle Vlasov distribution $f$ and the oscillation-center Vlasov distribution $\ov{F}$ is expressed in terms of the pull-back operator ${\sf T}_{\epsilon}$:
\begin{eqnarray}
f & = & {\sf T}_{\epsilon}\ov{F} \;\equiv\; \ov{F} \;+\; \epsilon\;{\sf G}_{1}\cdot\exd\ov{F} \;+\; \epsilon^{2} \left[ {\sf G}_{2}\cdot\exd\ov{F}
\;+\; \frac{1}{2}\;{\sf G}_{1}\cdot\exd\left( {\sf G}_{1}\cdot\exd\ov{F}\right) \right] \;+\; \cdots \nonumber \\
 & \equiv & f_{0} \;+\; \epsilon\;f_{1} \;+\; \epsilon^{2}\;f_{2} \;+\; \cdots,
\label{eq:FovF}
\end{eqnarray}
where the vectors $({\sf G}_{1}, {\sf G}_{2}, \cdots)$ are defined in Eqs.~(\ref{eq:G1_def}) and (\ref{eq:G2_def}).

\subsubsection{First-order pull-back operator.}
 
Since the oscillation-center Vlasov distribution $\ov{F}$ is independent of the fast wave space-time scales (i.e., it is independent of the eikonal phase $\Theta$), the eikonal-phase averaged part $\langle f_{1}\rangle$ of the first-order particle Vlasov distribution $f_{1} \equiv {\sf G}_{1}\cdot\exd
\ov{F}$ vanishes: $\langle f_{1}\rangle \equiv 0$ (since $\langle{\sf G}_{1}\rangle \equiv 0$). Its eikonal-phase dependent part $\wt{f}_{1} = 
\wt{f}_{10} + \epsilon_{0}\,\wt{f}_{11} + \cdots$, on the other hand, is expressed in terms of the lowest-order contribution
\begin{equation}
\wt{f}_{10} \;=\; i\;m\omega^{\prime}\;\wt{\vb{\xi}}\bdot\pd{\ov{F}}{{\bf p}} \;=\; -\;\frac{ie}{\omega^{\prime}}\; \left( \wt{{\bf E}}_{1} +
\frac{{\bf v}}{c}\btimes\wt{{\bf B}}_{1} \right)\bdot\pd{\ov{F}}{{\bf p}},
\label{eq:wtF_10}
\end{equation}
so that the zeroth velocity-moment is
\begin{equation}
\int\;\wt{f}_{10}\;d^{3}p \;=\; -i\int\;\ov{F}\;\pd{}{{\bf p}}\bdot\left(m\omega^{\prime}\,\wt{\vb{\xi}}\right)\;d^{3}\ov{p} \;=\; -\;\int\;\ov{F}\;\left(
i{\bf k}\bdot\wt{\vb{\xi}}\right)\;d^{3}\ov{p},
\label{eq:rho_10}
\end{equation}
from which we obtain the first-order polarization eikonal-amplitude
\begin{equation}
\wt{{\bf P}}_{1} \;\equiv\; \sum\;e\;\int\;\ov{F}\;\wt{\vb{\xi}}\;d^{3}\ov{p}.
\label{eq:P1_def}
\end{equation}
The first velocity-moment, on the other hand, is
\begin{eqnarray}
\int\;{\bf v}\;\wt{f}_{10}\;d^{3}p & = & -i\;\int\;\ov{F}\;\pd{}{\ov{{\bf p}}}\bdot\left(m\omega^{\prime}\,\wt{\vb{\xi}}\,\ov{{\bf v}}\right)\;d^{3}\ov{p} \;=\; 
-\;\int\;\ov{F}\;\left[ \left( i{\bf k}\bdot\wt{\vb{\xi}}\right)\;\ov{{\bf v}} \;+\; i\omega^{\prime}\;\wt{\vb{\xi}} \;\right]\;d^{3}\ov{p} \nonumber \\
 & = & \int\;\ov{F}\;\left[ -i\,\omega\;\wt{\vb{\xi}} \;+\; i{\bf k}\btimes\left(\wt{\vb{\xi}}\btimes\ov{{\bf v}} \right) \;\right] d^{3}\ov{p},
\label{eq:J_10}
\end{eqnarray}
where the first term in the last expression is associated with the first-order polarization current, while the second term is associated with the first-order magnetization eikonal-amplitude
\begin{equation}
\wt{{\bf M}}_{1} \;\equiv\; \sum\;e\;\int\;\ov{F}\;\wt{\vb{\xi}}\btimes\frac{\ov{{\bf v}}}{c}\;d^{3}\ov{p}.
\label{eq:M1_def}
\end{equation}
Note that only the moving electric-dipole contribution appears here since the intrinsic magnetization must be quadratic in $\vb{\xi}$.

\subsubsection{Second-order pull-back operator.}

The second-order particle Vlasov distribution $f_{2} \equiv {\sf G}_{2}\cdot\exd\ov{F} + \frac{1}{2}\;{\sf G}_{1}\cdot\exd( {\sf G}_{1}\cdot\exd\ov{F})$ has an eikonal-phase independent part expressed as
\begin{equation}
\langle f_{20}\rangle \;=\; \pd{}{{\bf p}}\bdot \left[\; \frac{e}{c} \left( \ov{{\bf A}}_{2} \;+\; {\rm Re}(\wt{{\bf B}}_{1}\btimes\wt{\vb{\xi}}^{*}) 
\right)\;\ov{F} \;+\; m^{2}\omega^{\prime\;2} \;{\rm Re}\left(\wt{\vb{\xi}}\;\wt{\vb{\xi}}^{*}\right)\bdot\pd{\ov{F}}{{\bf p}} \;\right],
\label{eq:F2_av}
\end{equation}
where only the lowest-order WBF terms are kept. Note that this second-order particle contribution has a vanishing zeroth velocity-moment
\begin{equation}
\int\; \langle f_{20}\rangle\;d^{3}p \;\equiv\; 0,
\label{eq:moment_0}
\end{equation}
since Eq.~(\ref{eq:F2_av}) is expressed as a momentum-space divergence. However, the first velocity-moment
\begin{eqnarray}
\int\;{\bf v}\;\langle f_{20}\rangle\; d^{3}p & = & -\; \int d^{3}\ov{p}\;\left[\; \ov{F} \left( \frac{e\ov{{\bf A}}_{2}}{mc} \;+\; \frac{e}{mc}\;
{\rm Re}\left(\wt{{\bf B}}_{1}\btimes\wt{\vb{\xi}}^{*}\right) \right) \;+\; m\omega^{\prime\;2}\;{\rm Re}(\wt{\vb{\xi}}\,\wt{\vb{\xi}}^{*})\bdot
\pd{\ov{F}}{\ov{{\bf p}}} \;\right] \nonumber \\
 & = & \int \ov{F} \left[\; -\;\frac{e}{mc}\,\ov{{\bf A}}_{2} \;+\; \omega^{\prime}\;{\bf k}\bdot
\left( \wt{\vb{\xi}}\;\wt{\vb{\xi}}^{*} + \wt{\vb{\xi}}^{*}\wt{\vb{\xi}} \right) \;\right] d^{3}\ov{p}
\label{eq:moment_1}
\end{eqnarray}
is non-vanishing, where we used the following identities in obtaining the last expression: $\partial_{{\bf p}}(m\omega^{\prime}\,\wt{\vb{\xi}}) \equiv 
\wt{\vb{\xi}}\,{\bf k}$ and
\[ \frac{e}{c}\;{\rm Re}\left(\wt{{\bf B}}_{1}\btimes\wt{\vb{\xi}}^{*}\right) \;\equiv\; m\omega^{\prime}\;\left[\; |\wt{\vb{\xi}}|^{2}\;{\bf k} \;-\; 
{\bf k}\bdot{\rm Re}(\wt{\vb{\xi}}^{*}\wt{\vb{\xi}}) \;\right], \]
which follows from the relation $(e/c)\,\wt{{\bf B}}_{1} = -\,m\omega^{\prime}\;{\bf k}\btimes\wt{\vb{\xi}}$ obtained from Eq.~(\ref{eq:xi_def}). Using the lowest-order second-order oscillation-center Hamiltonian (\ref{eq:H2_oc}), we obtain the identity
\begin{equation}
\int\;{\bf v}\;\langle f_{20}\rangle\; d^{3}p \;\equiv\; \int\;\ov{F}\;\left( -\,\frac{e}{mc}\,\ov{{\bf A}}_{2} \;+\; \pd{\ov{\Psi}_{20}}{\ov{{\bf p}}}
\right) \;d^{3}\ov{p}.
\label{eq:oc_identity}
\end{equation}
Lastly, the second-order kinetic energy
\begin{equation}
\int d^{3}p\;\left( \frac{|{\bf p}|^{2}}{2m}\right)\;\langle f_{20}\rangle \;\equiv\; \int\;\ov{F}\,\left( \ov{\Psi}_{20} \;-\; \frac{e}{c}\,\ov{{\bf A}}_{2}
\bdot\ov{{\bf v}} \right)\; d^{3}\ov{p}
\label{eq:FH_2}
\end{equation}
is naturally expressed in terms of the oscillation-center distribution $\ov{F}$ and the (lowest-order) second-order oscillation-center Hamiltonian 
(\ref{eq:H2_oc}).

\subsection{Ponderomotive Variational Principle}

The variational formulation for the exact Vlasov-Maxwell equations was presented in Sec.~\ref{sec:exact}. The covariance of the Vlasov part of the action functional has been used in Ref.~\cite{Brizard_2000b} to derive a variational formulation of the nonlinear gyrokinetic Vlasov-Maxwell equations, which describe the turbulent evolution of low-frequency electromagnetic fluctuations in a magnetized plasma with arbitrary geometry \cite{BH_07}. 

\subsubsection{Linear polarization and magnetization.}

The eikonal form of the first-order Maxwell equations can be expressed in terms of the eikonal-averaged second-order Lagrangian density 
\begin{eqnarray} 
\cal{L}_{20} & = & \frac{1}{4\pi} \left( |\wt{{\bf E}}_{1}|^{2} \;-\; |\wt{{\bf B}}_{1}|^{2} \right) \;-\;
\sum\; \int d^{3}\ov{p}\; \ov{F} \left[\; e\,\left( \ov{\Phi}_{2} \;-\; \frac{\ov{{\bf v}}}{c}\bdot\ov{{\bf A}}_{2} \right) \;+\; \ov{\Psi}_{20} \;\right] 
\nonumber \\
 & \equiv &  \frac{1}{4\pi}\;{\rm Re}\left( \wt{{\bf E}}_{1}\bdot\wt{{\bf D}}_{1}^{*} \;-\; \wt{{\bf B}}_{1}\bdot
\wt{{\bf H}}_{1}^{*} \right) \;-\; \sum\; e\;\int d^{3}\ov{p}\;\ov{F} \left( \ov{\Phi}_{2} \;-\; \frac{\ov{{\bf v}}}{c}\bdot\ov{{\bf A}}_{2} \right)
\label{eq:L20_Kchi}
\end{eqnarray}
where the last expression is obtained by writing the ponderomotive potential $\ov{\Psi}_{20}$ in terms of the first-order polarization and magnetization (\ref{eq:P1_def}) and (\ref{eq:M1_def}) as  
\[ \sum\; \int d^{3}\ov{p}\; \ov{F}\;\ov{\Psi}_{20} \;\equiv\; -\;{\rm Re}\left( \wt{{\bf P}}_{1}^{*}\bdot\wt{{\bf E}}_{1} \;+\; \wt{{\bf M}}_{1}^{*}\bdot
\wt{{\bf B}}_{1} \right), \]
which displays the standard relation between the ponderomotive Hamiltonian and the oscillation-center polarization and magnetization 
\cite{CK_chi,Kaufman_Kchi}.

We define the first-order eikonal-amplitudes for the macroscopic electromagnetic fields \cite{JDJ}
\begin{eqnarray}
\wt{{\bf D}}_{1} & \;\equiv\; & 4\pi\;\pd{\cal{L}_{20}}{\wt{{\bf E}}_{1}^{*}} \;\equiv\; 
\wt{{\bf E}}_{1} \;+\; 4\pi\,\wt{{\bf P}}_{1}, \label{eq:D1_def} \\
\wt{{\bf H}}_{1} & \equiv & -\;4\pi\;\pd{\cal{L}_{20}}{\wt{{\bf B}}_{1}^{*}} \;\equiv\; 
\wt{{\bf B}}_{1} \;-\; 4\pi\,\wt{{\bf M}}_{1}, \label{eq:H1_def}
\end{eqnarray}
in terms of the first-order eikonal-amplitudes for the polarization and magnetization (\ref{eq:P1_def}) and (\ref{eq:M1_def}). Hence, by using these definitions, the first-order Maxwell equations become
\begin{eqnarray}
0 & = & \pd{\cal{L}_{20}}{\wt{\Phi}_{1}^{*}} \;\equiv\; \pd{\wt{{\bf E}}_{1}^{*}}{\wt{\Phi}_{1}^{*}}\bdot
\pd{\cal{L}_{20}}{\wt{{\bf E}}_{1}^{*}} \;=\; i\,{\bf k}\bdot\frac{\wt{{\bf D}}_{1}}{4\pi}, \label{eq:D1_Poisson} \\
0 & = & \pd{\cal{L}_{20}}{\wt{{\bf A}}_{1}^{*}} \;\equiv\; \pd{\wt{{\bf E}}_{1}^{*}}{\wt{{\bf A}}_{1}^{*}}\bdot
\pd{\cal{L}_{20}}{\wt{{\bf E}}_{1}^{*}} \;+\; \pd{\wt{{\bf B}}_{1}^{*}}{\wt{{\bf A}}_{1}^{*}}\bdot\pd{\cal{L}_{20}}{\wt{{\bf B}}_{1}^{*}} 
\;=\; -\,i\,\frac{\omega}{c}\;\frac{\wt{{\bf D}}_{1}}{4\pi} \;-\; i\,{\bf k}\btimes\frac{\wt{{\bf H}}_{1}}{4\pi}, \label{eq:H1_Ampere}
\end{eqnarray}
where the eikonal amplitudes for the first-order charge-current densities are
\begin{eqnarray} 
\left( \begin{array}{c}
\wt{\rho}_{1} \\
\wt{{\bf J}}_{1}
\end{array} \right) & \equiv & \sum\; e\,\int d^{3}p\;\wt{f}_{10} \left( \begin{array}{c}
1 \\
{\bf v}
\end{array} \right) \;=\; \sum\; e\,\int d^{3}\ov{p}\;\ov{F} \left( \begin{array}{c}
-i\,{\bf k}\bdot\wt{\vb{\xi}} \\
-i\,\omega\;\wt{\vb{\xi}} \;+\; i\,{\bf k}\btimes\left( \wt{\vb{\xi}}\btimes\ov{{\bf v}}\right)
\end{array} \right) \nonumber \\
 & = & \left( \begin{array}{c}
-i\,{\bf k}\bdot\wt{{\bf P}}_{1} \\
-i\,\omega\;\wt{{\bf P}}_{1} \;+\; i\,{\bf k}c\btimes\wt{{\bf M}}_{1}
\end{array} \right), 
\label{eq:rhoJ_1}
\end{eqnarray}
which are consistent with Eqs.~(\ref{eq:P1_def}) and (\ref{eq:M1_def}). Note that Eq.~(\ref{eq:H1_Ampere}) is consistent with Eq.~(\ref{eq:D1_Poisson}), i.e., the dot product of Eq.~(\ref{eq:H1_Ampere}) with the wave vector ${\bf k}$ yields Eq.~(\ref{eq:D1_Poisson}).

\subsubsection{Nonlinear polarization and magnetization.}

The derivation of the lowest-order expressions for the eikonal-averaged second-order charge and current densities, and the eikonal corrections corresponding to the ponderomotive polarization and magnetization, proceeds with the introduction of the second-order eikonal-averaged Vlasov-Maxwell action functional
\begin{eqnarray}
\cal{L}_{2} & = & \left. \left. \frac{1}{4\pi} \right[ {\rm Re}\left( \wt{{\bf E}}_{1}\bdot\wt{{\bf D}}_{1}^{*} \;-\; \wt{{\bf B}}_{1}\bdot
\wt{{\bf H}}_{1}^{*} \right) \;+\; \epsilon_{0} \left( {\bf E}_{0}\bdot\ov{{\bf D}}_{2} \;-\; {\bf B}_{0}\bdot\ov{{\bf H}}_{2} \right) \right] \nonumber \\
 &  &-\;\sum\; e\;\int d^{3}\ov{p}\;\ov{F} \left( \ov{\Phi}_{2} \;-\; \frac{\ov{{\bf v}}}{c}\bdot\ov{{\bf A}}_{2} \right).
\label{eq:action_2}
\end{eqnarray}
We first note that, since the second-order eikonal-averaged Vlasov Lagrangian density (\ref{eq:action_2}) is independent of the background scalar potential $\Phi_{0}$, we find
\begin{equation}
\ov{\rho}_{20} \;\equiv\; -\;\pd{\cal{L}_{2}}{\Phi_{0}} \;=\; 0,
\label{eq:rho_Phi}
\end{equation}
and, thus, the second-order oscillation-center eikonal-averaged charge density must be generated by ponderomotive polarization effects [see 
Eqs.~(\ref{eq:rho2_1})-(\ref{eq:J2_1})] to first order (in $\epsilon_{0}$) in eikonal analysis. This result is consistent with the fact that the zeroth-moment (\ref{eq:moment_0}) of the second-order eikonal-averaged particle Vlasov distribution $\langle f_{20}\rangle$ vanishes.

The second-order ponderomotive (eikonal-averaged) polarization and magnetization are obtained from the action functional (\ref{eq:action_2}) as
\begin{eqnarray}
\ov{{\bf P}}_{2} & = & \epsilon_{0}^{-1}\pd{\cal{L}_{2}}{{\bf E}_{0}} - \frac{\ov{{\bf E}}_{2}}{4\pi} \;=\; \sum\;
\int\; \ov{\vb{\pi}}_{2}\;\ov{F}\;d^{3}\ov{p}, \label{eq:P2_def} \\
\ov{{\bf M}}_{2} & = & \epsilon_{0}^{-1}\pd{\cal{L}_{2}}{{\bf B}_{0}} + \frac{\ov{{\bf B}}_{2}}{4\pi} \;=\; \sum\;\int 
\left( \ov{\vb{\mu}}_{2} \;+\; \ov{\vb{\pi}}_{2}\btimes\frac{\ov{{\bf v}}}{c} \right)\;\ov{F}\;d^{3}\ov{p},
\label{eq:M2_def}
\end{eqnarray}
where $\ov{\vb{\pi}}_{2}$ and $\ov{\vb{\mu}}_{2}$ are defined in Eqs.~(\ref{eq:pi_2}) and (\ref{eq:mu_2}). The second-order ponderomotive charge and current densities are 
\begin{eqnarray}
\ov{\rho}_{21} & = & -\,\nabla\bdot\ov{{\bf P}}_{2}, \label{eq:rho2_1} \\
\ov{{\bf J}}_{21} & = & \pd{\ov{{\bf P}}_{2}}{t} \;+\; c\,\nabla\btimes\ov{{\bf M}}_{2}, 
\label{eq:J2_1}
\end{eqnarray}
respectively. 

Lastly, the second-order eikonal-averaged Maxwell equations can also be written as
\begin{equation}
\nabla\bdot\ov{{\bf D}}_{2} \;=\; 0 \;\;\;{\rm and}\;\;\; \nabla\btimes\ov{{\bf H}}_{2} \;-\; \frac{1}{c}\,
\pd{\ov{{\bf D}}_{2}}{t} \;=\; \frac{4\pi}{c}\;\ov{{\bf J}}_{20},
\label{eq:Maxwell_2_HD}
\end{equation}
which requires $\ov{{\bf J}}_{20}$ to be divergenceless \cite{Cary_Kaufman}, as is also required for the eikonal-averaged second-order charge and current densities $\ov{\rho}_{2} = \ov{\rho}_{20} + \epsilon_{0}\,\ov{\rho}_{21}$ and $\ov{{\bf J}}_{2} = 
\ov{{\bf J}}_{20} + \epsilon_{0}\,\ov{{\bf J}}_{21}$ to satisfy the eikonal-averaged second-order charge conservation law
\begin{equation}
\pd{\ov{\rho}_{2}}{t} \;+\; \nabla\bdot\ov{{\bf J}}_{2} \;=\; 0.
\label{eq:charge_2}
\end{equation}
Hence, the condition (\ref{eq:rho_Phi}) immediately implies that $\nabla\bdot\ov{{\bf J}}_{20} \equiv 0$ \cite{Cary_Kaufman}, while the second-order ponderomotive charge and current densities (\ref{eq:rho2_1})-(\ref{eq:J2_1}) explicitly satisfy Eq.~(\ref{eq:charge_2}).

We have, thus, achieved the purpose of this Section, which was to show how the ponderomotive polarization and magnetization (\ref{eq:PM2_push}) can be self-consistently derived from a variational principle (\ref{eq:action_2}) for the oscillation-center Vlasov-Maxwell equations. 
We note that ponderomotive polarization and magnetization effects in magnetized plasmas have also been investigated in Refs.~\cite{SK,SKD} by using a Low-Lagrangian-type variational formulation. 

\subsection{\label{subsec:pond_push}Ponderomotive Push-forward Derivation}

We now proceed with the complementary derivation of the ponderomotive polarization and magnetization by the ponderomotive push-forward method. 
By inserting the oscillation-center transformation generated by Eqs.~(\ref{eq:G1_def}) and (\ref{eq:G2_def}) into the reduced displacement 
(\ref{eq:rhoepsilon_def}), we obtain the oscillation-center displacement
\begin{equation} 
\vb{\rho}_{\epsilon} \;\equiv\; \epsilon\,\vb{\xi} \;+\; \frac{\epsilon^{2}}{2} \left( \vb{\xi}\bdot\nabla\vb{\xi}
\;+\; m\;\frac{d\vb{\xi}}{dt}\bdot\pd{\vb{\xi}}{{\bf p}} \right) \;-\; \epsilon^{2}\,G_{2}^{{\bf x}} \;+\; \cdots,
\label{eq:rho_oc}
\end{equation}
where Eq.~(\ref{eq:dxi_dt}) was used and the first-order WBF corrections are omitted in what follows. While the second-order term $G_{2}^{{\bf x}} \equiv -\,\partial_{{\bf p}}S_{2}$ appears on the right side of Eq.~(\ref{eq:rho_oc}), it does not contribute to the relations derived below because its contributions are of higher order in $\epsilon$.

\subsubsection{Linear polarization and magnetization.}

The linear displacement $\vb{\rho}_{\epsilon} = \epsilon\,\vb{\xi}$ is eikonal-phase dependent and its amplitude is
$\wt{\vb{\rho}}_{\epsilon} = \epsilon\,\wt{\vb{\xi}}$. Hence, the push-forward formula (\ref{eq:Pol_push}) yields the linear polarization 
\begin{equation} 
\wt{{\bf P}}_{1} \;=\; \sum\,e\,\ov{n}\;\ov{\left[\,\wt{\vb{\xi}}\,\right]} \;=\; -\;\sum\;\ov{n}\;\pd{\ov{[\ov{\Psi}_{20}]}}{\wt{{\bf E}}_{1}^{*}}.
\label{eq:P1_pf}
\end{equation}

The push-forward formula (\ref{eq:J_push}), on the other hand, yields the linear expression for the eikonal-dependent current
\[ \wt{{\bf J}}_{1} \;=\; -\,i\omega\;\wt{{\bf P}}_{1} \;+\; i\,{\bf k}\btimes \left( \sum\,e\,\ov{n}\;\ov{\left[\,
\wt{\vb{\xi}}\btimes\ov{{\bf v}} \,\right]} \right) \;=\; -\,i\omega\;\wt{{\bf P}}_{1} \;+\; i\,{\bf k}c\btimes\wt{{\bf M}}_{1} \]
so that the linear magnetization is
\begin{equation} 
\wt{{\bf M}}_{1} \;=\; \sum\,e\,\ov{n}\;\ov{\left[\,\wt{\vb{\xi}}\btimes\frac{\ov{{\bf v}}}{c}\,\right]} \;=\; -\;\sum\;\ov{n}\;\pd{\ov{[\ov{\Psi}_{20}]}}{\wt{{\bf B}}_{1}^{*}},
\label{eq:M1_pf}
\end{equation}
where we used the lowest-order expression $\dot{\ov{{\bf x}}}_{\epsilon} = \ov{{\bf v}}$ for the reduced particle velocity. Note that the first-order magnetization (\ref{eq:M1_pf}) only has a moving electric-dipole contribution (i.e., $\wt{\vb{\mu}}_{1} \equiv 0$) since the intrinsic magnetization (\ref{eq:mu_def}) is quadratic in $\vb{\rho}_{\epsilon}$.

\subsubsection{Ponderomotive polarization and magnetization.}

Ponderomotive polarization and magnetization effects enter at second order through eikonal-phase averaging. We begin with the eikonal-averaged reduced displacement
\[ \langle\vb{\rho}_{\epsilon}\rangle \;=\; \epsilon^{2}\;{\rm Re}\left[\; i\,{\bf k}\bdot\left(\wt{\vb{\xi}}^{*}\,\wt{\vb{\xi}}\right) \;-\; i\,m
\omega^{\prime}\wt{\vb{\xi}}\bdot\pd{\wt{\vb{\xi}}^{*}}{{\bf p}} \;\right] \;=\; \epsilon^{2}\;{\bf k}\btimes\left( i\,\wt{\vb{\xi}}\btimes
\wt{\vb{\xi}}^{*}\right) \;\equiv\; \epsilon^{2}\;\left( e^{-1}\,\ov{\vb{\pi}}_{2} \right), \]
from which we recover the ponderomotive electric-dipole moment (\ref{eq:pi_2}). Hence, the push-forward relation (\ref{eq:Pol_push}) yields the second-order ponderomotive polarization
\begin{equation} 
\ov{{\bf P}}_{2} \;=\; \epsilon^{-2}\,\sum\;e\,\ov{n}\;\ov{[\langle\vb{\rho}_{\epsilon}\rangle]} \;=\; 
\sum\;\ov{n}\;\ov{[\ov{\vb{\pi}}_{2}]} \;\equiv\; -\;\sum\;\ov{n}\;\pd{\ov{[\ov{\Psi}_{21}]}}{{\bf E}_{0}}.
\label{eq:P2_pf}
\end{equation}
We note that the ponderomotive magnetization contribution to the ponderomotive polarization is obtained from the relativistic correction 
(\ref{eq:pol_rel}) as
\begin{eqnarray}
e\;\left\langle \vb{\rho}_{\epsilon}\;{\sf T}_{\epsilon}^{-1}(\gamma - 1)\right\rangle & = & \frac{e}{c^{2}}\;\left\langle \vb{\rho}_{\epsilon}\;
\left( \ov{{\bf v}}\bdot\frac{d_{\epsilon}\vb{\rho}_{\epsilon}}{dt}\right)\right\rangle \;+\; \cdots \nonumber \\
 & = & \epsilon^{2} \left[\; \frac{e}{c^{2}}\;\ov{{\bf v}}\btimes\left( i\,\omega^{\prime}\;\wt{\vb{\xi}}\btimes\wt{\vb{\xi}}^{*}\right) \;\right] \;\equiv\; 
\frac{\ov{{\bf v}}}{c}\btimes\left( \epsilon^{2}\;\ov{\vb{\mu}}_{2}\right),
\label{eq:P2_mu}
\end{eqnarray}
where $d_{\epsilon}\ov{{\bf x}}/dt \equiv \ov{{\bf v}}$ (to lowest order in $\epsilon$) and we have omitted the relativistic correction 
$(|\ov{{\bf v}}|^{2}/2c^{2})\langle\vb{\rho}_{\epsilon}\rangle$ to the ponderomotive polarization (\ref{eq:P2_pf}). Furthermore, we note that the quadrupolar contribution to the ponderomotive polarization (\ref{eq:P2_pf}) is of the form $\epsilon^{2}\,\nabla\bdot\ov{{\sf Q}}_{2}$, where the ponderomotive electric quadrupole moment (tensor) is defined as $\ov{{\sf Q}}_{2} \equiv \sum\,e\ov{n}\;{\rm Re}([\wt{\vb{\xi}}\,\wt{\vb{\xi}}^{*}])$, which is one order higher than the ponderomotive polarization in the eikonal analysis (i.e., $|\nabla\bdot\ov{{\sf Q}}_{2}| \sim \epsilon_{0}\,
|\ov{{\bf P}}_{2}|$) and ponderomotive quadrupolar polarization is therefore omitted.

Next, we derive an expression for the second-order ponderomotive magnetization. First, the moving-dipole contribution is
$\langle(e\,\vb{\rho}_{\epsilon}\btimes\dot{\ov{{\bf x}}}_{\epsilon})\rangle = e\,\langle\vb{\rho}_{\epsilon}\rangle
\btimes\dot{\ov{{\bf x}}}_{\epsilon} = \epsilon^{2}\;\ov{\vb{\pi}}_{2}\btimes\ov{{\bf v}}$ to lowest order in $\epsilon$. Second, the intrinsic ponderomotive magnetization is
\[ \left\langle\left(\frac{e}{2}\,\vb{\rho}_{\epsilon}\btimes\frac{d_{\epsilon}\vb{\rho}_{\epsilon}}{dt}\right)\right\rangle \;=\; 
e\,\epsilon^{2}\;\omega^{\prime}\;\left( i\,\wt{\vb{\xi}}\btimes\wt{\vb{\xi}}^{*}\right) \;\equiv\; \epsilon^{2}\;
c\,\ov{\vb{\mu}}_{2}, \]
where we used $(\vb{\rho}_{\epsilon}, d_{\epsilon}\vb{\rho}_{\epsilon}/dt) \rightarrow (\epsilon\,\wt{\vb{\xi}},\; -i\epsilon\,\omega^{\prime}
\wt{\vb{\xi}})$. The push-forward relation (\ref{eq:Mepsilon_def}) yields the second-order ponderomotive magnetization
\begin{eqnarray} 
\ov{{\bf M}}_{2} & = & \epsilon^{-2}\,\sum\;\frac{e}{c}\,\ov{n}\;\ov{\left[\left\langle\vb{\rho}_{\epsilon}\btimes\left( \frac{1}{2}\;\frac{d_{\epsilon}\vb{\rho}_{\epsilon}}{dt} \;+\; \frac{d_{\epsilon}\ov{{\bf x}}}{dt} \right)\right\rangle\right]} \;\equiv\; \sum\;\ov{n}\;\ov{\left[ 
\ov{\vb{\mu}}_{2} \;+\; \ov{\vb{\pi}}_{2}\btimes \frac{\ov{{\bf v}}}{c} \right]} \nonumber \\
 & \equiv & -\;\sum\;\ov{n}\;\pd{\ov{[\ov{\Psi}_{21}]}}{{\bf B}_{0}}.
\label{eq:M2_pf}
\end{eqnarray}
Equations (\ref{eq:P2_pf}) and (\ref{eq:M2_pf}) represent the main results of this Section. They exhibit the complementarity of the push-forward and variational methods for the case of the oscillation-center dynamics of charged particles in a weakly-magnetized background plasma.

\section{\label{sec:sum}Summary}

From a historical perspective, the derivation of oscillation-center Hamiltonian dynamics provided an ideal battleground for the application of Lie-transform perturbation methods in plasma physics, in which the Berkeley School of Plasma Physics, led by Allan N.~Kaufman and his collaborators, played a fundamental role. In the spirit of the Berkeley School, the present paper demonstrated the interconnectedness of the Lie-transform perturbation method with the variational approach.

The variational formulation (\ref{eq:action_2}) of the {\it oscillation-center} Vlasov-Maxwell equations provides a simple description of self-consistent ponderomotive polarization and magnetization effects in weakly-magnetized plasmas, derived by variational derivatives with respect to the background electromagnetic fields $({\bf E}_{0},{\bf B}_{0})$. Although the treatment presented here does not include relativistic effects, the variational formalism is naturally suitable for relativistic generalization.

Ponderomotive effects for unmagnetized plasmas \cite{Cary_Kaufman} can be easily recovered by setting ${\bf A}_{0} \equiv 0$ in the final expressions. Ponderomotive effects in strongly-magnetized plasma, on the other hand, require a two-step transformation: first, a transformation from particle to guiding-center phase space is introduced \cite{RGL_83} to be followed by a transformation to oscillation-center phase space \cite{GKL,Cary_Kaufman}. Lastly, low-frequency ponderomotive polarization and magnetization effects in magnetized plasmas are described in terms of the nonlinear gyrokinetic Vlasov-Maxwell equations (e.g., see Ref.~\cite{BH_07}) or the nonlinear electromagnetic drift-fluid equations \cite{Brizard_NFLR}.

\acknowledgments

Portions of this paper were presented as an invited talk in celebration of Allan Kaufman's 80$^{{\rm th}}$ birthday during the KaufmanFest 2007 Symposium in Berkeley, October 6-7, 2007. The Symposium also marked the 50$^{{\rm th}}$ anniversary of Allan's first paper in plasma physics (with S.~Chandrasekhar and K.~M.~Watson \cite{CKW}).

Much of the work presented in this paper was done by the Author with support from the U.~S.~Department of Energy through the Offices of Fusion Energy Sciences and Basic Energy Sciences as well as the National Science Foundation. 

Last but not least, Allan's constant support and encouragement over the past (nearly) 20 years are most gratefully acknowledged. I have learned greatly from working with him and have benefited immensely from his keen physical insights. In fact, the present manuscript benefited greatly from Allan's critical reading and insightful suggestions. I am proud to declare myself a member of the Berkeley School of Plasma Physics.

\end{document}